\documentclass[preprint,12pt]{elsarticle}
\usepackage{graphics}
\usepackage{graphicx}
\usepackage{amsmath}
\usepackage{amssymb}
\usepackage{longtable}
\usepackage[tight]{subfigure}
\journal{International Journal of Mass Spectrometry}

\begin{document}
\begin{frontmatter}



\title{Verifying the accuracy of the TITAN Penning-trap mass spectrometer}

\author{M. Brodeur$^{1,2}$\corref{cor1}}
\ead{brodeur@nscl.msu.edu}
\author{V.L.~Ryjkov$^{1}$, T. Brunner$^{1,3}$, S. Ettenauer$^{1,2}$, A. T. Gallant$^{1,2}$,  V.V.~Simon$^{1,4,5}$, M.J.~Smith$^{1,2}$}
\author{A. Lapierre$^{1}$\corref{cor1}} 
\author{R. Ringle$^{1}$\corref{cor1}}
\author{P. Delheij$^{1}$, M. Good$^{1}$, D. Lunney$^{6}$ and J. Dilling$^{1,2}$}
\cortext[cor1]{Present address: National Superconducting Cyclotron Laboratory, Michigan State University, East Lansing, MI, 48824, USA}
\address{$^{1}$ TRIUMF, 4004 Wesbrook Mall, Vancouver, BC, V6T 2A3, Canada}
\address{$^{2}$ Department of Physics and Astronomy, University of British Columbia, Vancouver, BC, V6T 1Z1, Canada}
\address{$^{3}$ Physik Department E12, Technische Universit\"{a}t M\"{u}nchen, James Franck Str., D-85748 Garching, Germany}
\address{$^{4}$ Max-Planck-Institut f\"{u}r Kernphysik, Saupfercheckweg 1, D-69117 Heidelberg, Germany}
\address{$^{5}$ Physikalisches Institut, Ruprecht-Karls-Universit\"{a}t Heidelberg, Philosophenweg 12, 69120 Heidelberg, Germany}
\address{$^{6}$CSNSM-IN2P3-CNRS, Universit\'{e} Paris 11, 91405 Orsay, France}

\begin{abstract}
TITAN (TRIUMF's Ion Traps for Atomic and Nuclear science) is an online facility designed to carry out high-precision mass measurements on singly and highly charged radioactive ions. The TITAN Penning trap has been built and optimized in order to perform such measurements with an accuracy in the sub ppb-range. A detailed characterization of the TITAN Penning trap is presented and a new compensation method is derived and demonstrated, verifying the performance in the range of sub-ppb.
\end{abstract}

\end{frontmatter}


\section{Introduction} \label{Sec:1}

Penning traps have proven to be the most precise devices for mass spectrometry, both  for stable and unstable isotopes \cite{Bla06}. For unstable isotopes with half-lives between 5 ms and a few seconds, performing such measurements is a challenging task because the unstable isotopes produced via nuclear reactions need to be delivered in a fast and efficient manner while the subsequent measurement need to reach the desired precision. However, the study of  several phenomena far from stability benefits from a precise mass determination. These physical processes includes: change in the nuclear structure (\cite{Kan09}-\cite{Hak08}), the determination of the exact path of the -r, -rp and -$\nu$p processes (\cite{Sch07}-\cite{Web08}), the improvement of the halo nuclei charge radius precision (\cite{Mue07}-\cite{Rin09a}) and the testing of the CVC hypothesis (\cite{Har10} and references therein). The needed relative uncertainty on the mass determination for these various cases varies from $\delta m/m \sim$ 10$^{-6}$ to 10$^{-8}$. To reach such precision while being accurate, great care need to be taken on identifying and minimizing the various sources of systematic errors on the mass determination.

There are currently several experiments dedicated to precise mass measurements of short-lived nuclei including ISOLTRAP \cite{Muk08} at ISOLDE/CERN, CPT \cite{Cla03} at ATLAS/ANL, SHIPTRAP \cite{Blo07} at GSI, LEBIT \cite{Rin09b} at NSCL/MSU, JYFLTRAP \cite{Elo08} at JYFL, TRIGA-TRAP at TRIGA Mainz \cite{Ket08} and TITAN \cite{Dil03, Dil06} at ISAC/TRIUMF. These experiments are complementary since they are set-up at different production facilities and all have a specific reach and access to isotopes. TITAN as for example succeeded in measuring several masses in the light mass region, including $^{8}$He \cite{Ryj08}, $^{6}$Li \cite{Bro09}, $^{8,9,11}$Li \cite{Smi08}, $^{9,10,11}$Be \cite{Rin09a} and $^{12}$Be \cite{Ett10}. 

Although the TITAN mass measurement program thrived in measuring the masses of very-short-lived (as low as 8.8 ms for $^{11}$Li) halo nuclei, the possibility of performing high-precision mass measurements on highly charged unstable ions (HCI) is a distinctive feature of TITAN. As it was previously demonstrated with stable species at the SMILETRAP experiment \cite{Ber02}, HCIs are used because the precision of mass measurements performed using Penning traps linearly increases with the charge state \cite{Dil06}. 

Over the past 30 years, extensive work has been done to identify the various factors that limits the precision and accuracy of Penning trap mass spectrometry. These sources of systematic errors includes: magnetic field inhomogeneities, misalignment with the magnetic field, harmonic distortion and anharmonicities of the trapping potential, temporal fluctuations of the magnetic field, relativistic effects and ion-ion interactions. These effects are described in the literature (\cite{Bro82}-\cite{Bol92}) and studied for specific Penning trap spectrometers \cite{Kel03, Ber02, Van06}. In this paper we describe how these systematic effects where minimized for the TITAN Penning trap and we give an estimate for their upper values. Then we determined experimentally the so-called mass-dependant systematic shift of the measured frequency ratio which is a combination of all the effects that are found to depend linearly on the difference in mass-to-charge ratio between the calibrant and ion of interest.

\section{TOF-ICR Penning trap mass measurements basics} \label{Sec:2}

The basic principal behind Penning-trap mass spectrometry \cite{Gar78} consists of measuring the cyclotron frequency 
\begin{equation}\label{eq:1}
\nu_{c} = \frac{1}{2 \pi}\frac{q B}{M}
\end{equation} 
of an ion of mass $M$ and charge $q$ in a magnetic field $B$. Knowing the field strength (this requires a reference mass, as discussed below) and charge state of the ion, one can then obtain its mass. In order to reach a high precision, on the order of $\delta m/m \leq$ 5$\times$10$^{-9}$ on the ion's mass, several requirements need to be met. Principally, a magnetic field that is homogenous in the region the ion is stored and a sufficiently long observation time are needed. These requirements are more easily fulfilled by confining the trapped ions in a small volume. 
\begin{figure}[ht]
  \begin{center}{
    \includegraphics[width=0.6\textwidth]{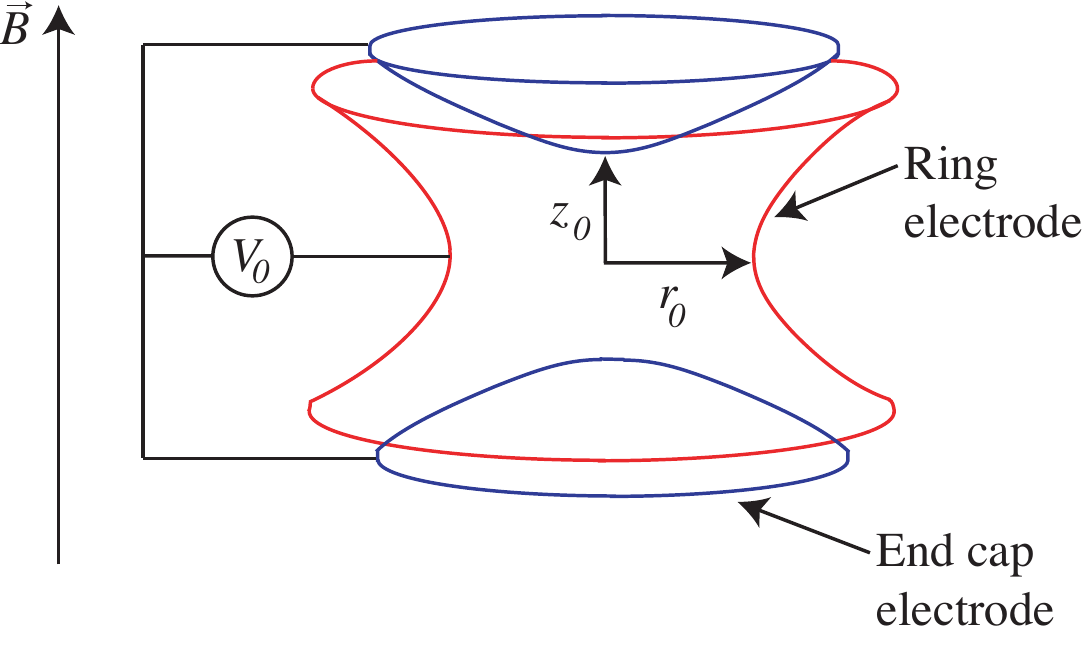}}
    \caption{(Colour on-line) Schematic diagram of a Penning trap including: the hyperboloids forming the ring electrode (red), the hyperboloid forming the end caps (blue) and an axial magnetic field.}
    {\label{fig:hyperPenn}}
  \end{center}
\end{figure}

The Penning trap \cite{Deh67} (schematic shown in figure~\ref{fig:hyperPenn}) is a type of ion trap that achieves this using a strong homogenous magnetic field overlaid with a quadrupolar electrostatic potential. This potential is created by applying a potential difference $V_{0}$ between a set of electrodes, typically orthogonal hyperboloids of revolution: one forming the ring electrode and the other forming the end cap electrodes. The resulting potential in such a configuration is given by:
\begin{equation}\label{eq:2}
V_{2}(z,r) = \frac{V_{0}}{2 d_{0}^{2}}(z^{2} - r^{2}/2),
\end{equation} 
where ($z$, $r$) are the axial and radial coordinates and \(d_{0} = \sqrt{z_{0}^{2}/2 + r_{0}^{2}/4}\) is typically defined as the characteristic length of the trap. The parameters ($z_{0}$, $r_{0}$) are the distances from the trap centre to the electrodes as defined in figure~\ref{fig:hyperPenn}. This electrostatic potential configuration axially traps the ions, while the radial confinement is provided by a magnetic field $B$ parallel to the end cap axis. Note that similar confinement properties can also be achieved using a cylindrical sets of electrodes that have been orthogonalized \cite{Gab84} instead of the hyperbolical ones. 

There exist analytical solutions for the ion motion in a Penning trap that are extensively studied in the literature (see for example \cite{Bro86} and \cite{Kre92}). The ion motion is composed of three eigenmotions: one axial of frequency \(\nu_{z} = \sqrt{qV_{0}/(M d_{0}^{2})}/(2 \pi)\) and two radial of frequencies \(\nu_{\pm} = \nu_{c}/2 \pm \sqrt{\nu_{c}^{2}/4 -  \nu_{z}^{2} / 2} \), called reduced cyclotron $(\nu_{+})$ and magnetron $(\nu_{-})$. The three eigenfrequencies typically have the following hierarchy: \(\nu_{+} \gg \nu_{z} \gg \nu_{-}\) and it can be noted that the cyclotron frequency is not an eigenfrequency, but for an ideal (purely quadrupole) trap the relation
\begin{equation}\label{eq:3}
\nu_{c} =  \nu_{+} + \nu_{-}
\end{equation} 
holds. 

Penning-trap mass spectrometers that uses the time-of-flight ion-cyclotron resonance (TOF-ICR) technique \cite{Gra80} make use of equation~\eqref{eq:3} in order to obtain the cyclotron frequency of the ion and ultimately, using equation~\eqref{eq:1}, its mass. It should be stressed that equation~\eqref{eq:3} is only valid for an ideal trap. This paper will present estimates of the various sources of deviations from the ideal trap and their effects on the measured cyclotron frequency, that ultimately affects the accuracy on a mass measurement in this case, at TITAN. 

At TITAN, the TOF-ICR technique is used to measure the cyclotron frequency. In this technique, the ion's eigenmotions are excited by applying a radio-frequency (RF) field. Two types of excitations are typically used: dipole and quadrupole. The application of the first one in the radial plane at either $\nu_{-}$ or $\nu_{+}$ results in the excitation of the corresponding eigenmotion. The application of a quadrupolar excitation at the sum frequency \(\nu_{-} + \nu_{+}\) results in a beating between the two different modes. Therefore, by the application of an RF excitation at the frequency $\nu_{q}$ on an ion initially in a pure magnetron motion, a complete conversion of the ion's motion into a pure reduced cyclotron motion will occur when \(\nu_{q} = \nu_{-} + \nu_{+}\). A full conversion only happens when the RF excitation amplitude $V_{q}$ and time $T_{q}$ are related by \cite{Kon95}:
\begin{equation}\label{eq:3b}
V_{q} = \frac{2 \pi B a^{2} \eta_{0}}{T_{q}},
\end{equation} 
where $a$ is the distance from the trap centre at which the RF field amplitude equals $V_{q}$ and $\eta_{0}$ = 1, 3, 5, ... are integer values of the conversion factor $\eta$ that allows a full conversion.

Because \(\nu_{+} \gg \nu_{-}\), a conversion will result in a gain of the ion's kinetic energy $E_{r}$ in the radial plane. The ion's cyclotron frequency can be derived by determining the excitation frequency that yields the largest increase in the ion's kinetic energy. This increase is found by releasing the ion from the trap and measuring its flight time taken to reach a detector situated outside the strong magnetic field region. On the way to the detector, the interaction of the ion's motion magnetic dipole moment $\mu$ (resulting from the RF excitation) with the magnetic field gradient $\partial B_{z} / \partial z$ induces a force \(\vec F  = \vec \nabla (\vec \mu \cdot \vec B) = - \frac{E_{r}}{B_{0}} \frac{\partial B_{z}}{\partial z} \hat{z}\) that axially accelerates the ions. Since \(F \propto E_{r}\), the acceleration is the greatest when \(\nu_{q} = \nu_{-} + \nu_{+}\), yielding a shorter time of flight. The time of flight as a function of the excitation frequency is described analytically by the following integral \cite{Kon95}:
\begin{equation}\label{eq:4}
\mbox{TOF}(\nu_{q}) = \int_{z_{0}}^{z_{1}}{ \left\{ \frac{M}{2 \cdot \left [ E_{0} - q \cdot V(z) - \mu(\nu_{q}) \cdot B(z) \right ]} \right \}^{1/2} dz},
\end{equation}
where $z_{0}$ and $z_{1}$ are the initial and final position of the ion prior and after ejection, $E_{0}$ is its kinetic energy upon leaving the trap, $V(z)$ and $B(z)$ are the electrical potential and magnetic fields along the path of the ion.
\begin{figure}
  \begin{center}
    \includegraphics[width=0.6\textwidth]{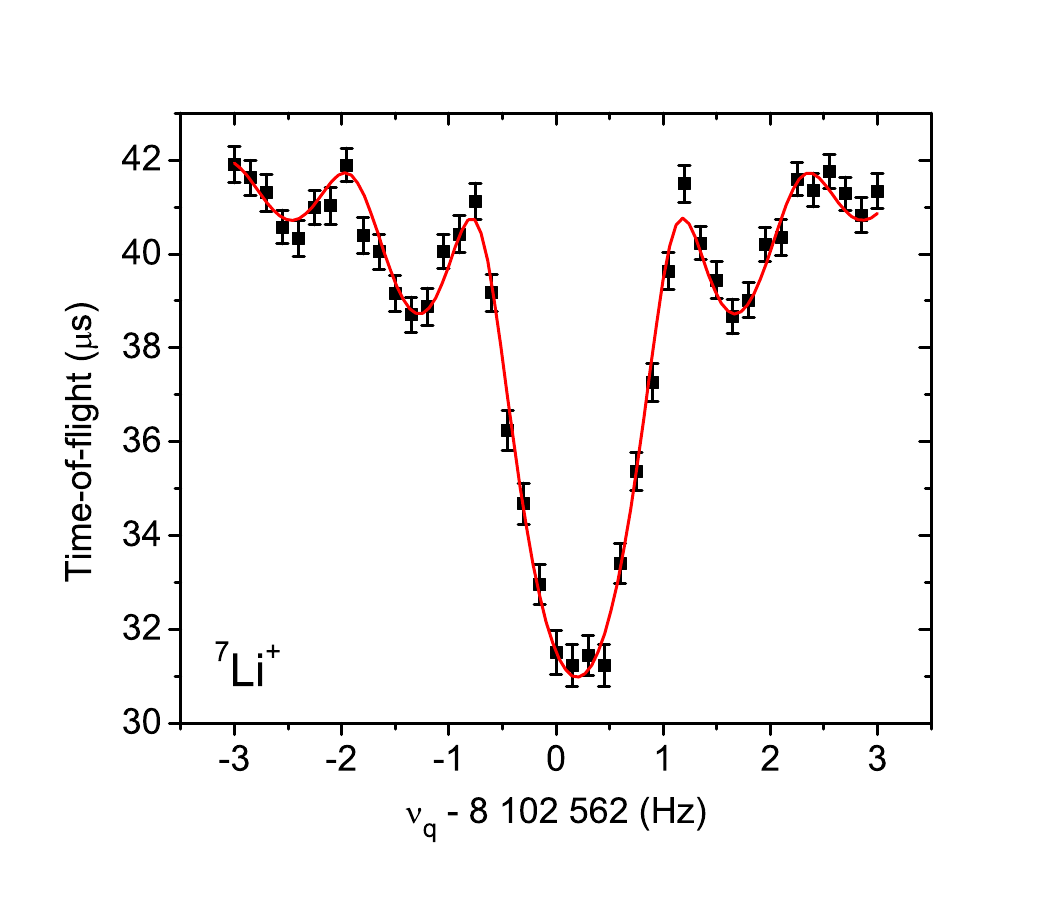}
    \caption{$^{7}$Li$^{+}$ cyclotron frequency resonance taken with a 900 ms excitation time. The solid line is a fit of the theoretical line shape \cite{Kon95} to the data.}{\label{fig:Li7TOF}}
  \end{center}
\end{figure}
Repeating the injection, excitation, extraction, and time-of-flight measurement process for different frequencies produces a time-of-flight spectrum such as the one in figure~\ref{fig:Li7TOF} from which the cyclotron frequency is derived as the centroid of the minimum. 

To measure the magnetic field of the Penning trap a measurement of the cyclotron frequency of calibrant ions has to be carried out. Typically the calibrant mass is more precisely known than the mass of the ion of interest. Hence, the measured quantity from which the mass is computed is the frequency ratio 
\begin{equation}\label{eq:5}
R = \frac{\nu_{c,cal}}{\nu_{c}} = \frac{q_{cal} \cdot M}{q \cdot M_{cal}}.
\end{equation}  
Note that the calibrant ion cyclotron frequency $\nu_{c,calib.}$ value at the time of the $\nu_{c}$ measurement is approximated by a linear interpolation of two calibration measurements enclosing the measurement of the ion of interest. 

In comparative mass spectrometry, the quantity of interest is the atomic mass of a neutral atom, which is given by:
\begin{equation}\label{eq:6}
m = \frac{q}{q_{cal}} \cdot \overline{R} \cdot (m_{cal} -
q_{cal} \cdot m_{e} + B_{e,cal}) + q \cdot m_{e} - B_{e},
\end{equation}
where $\overline{R}$ is the weighted mean of all measured frequency ratios, $B_{e,cal}$ and $B_{e}$ are the calibrant's and ion of interest's electron binding energies, $q_{cal}$ and $q$ are their respective charge states, $m_{e}$ is the electron mass and $m_{cal}$ is the calibrant atomic mass. The statistical uncertainty on a mass measurement is given by the following relation \cite{Bol01, Rin09b}:
\begin{equation}\label{eq:7}
\frac{\delta m}{m} = \frac{\delta \nu_{c}}{\nu_{c}} =  \frac{\gamma \cdot m}{q \cdot B \cdot T_{q} \cdot \sqrt{N_{ion}}},
\end{equation}
where $N_{ion}$ is the total number of ions detected for the measurement and $\gamma$ is an experiment-dependent constant given by the quality factor \cite{Kel03,Rin09b} of the time-of-flight resonance spectra.

Equation~\eqref{eq:7} describes the precision on a mass measurement using the TOF-ICR technique. However, ultimately the accuracy is limited by systematic errors that arise from a number of factors. For instance, the trap electrodes do not extend to infinity and are truncated. Also, holes in the two end-cap electrodes are required to inject and extract the ions from the trap and hence disturb the ideal potential. Moreover, the ideal trap assumes perfect geometrical alignment of all applied electrostatic and magnetic fields. In reality, misalignments between each trap electrode and distortion in the shape of the electrodes exist due to technical limitations in the achievable machining tolerances and affect the trapping potential \cite{Bro86,Bol90}. There are also misalignments of the trap's electrode structure principal axis with the magnetic field axis, and deviations of the magnetic field in the trapping region. Other effects arise due to the Coulomb interaction between stored ions. Moreover, fluctuations of the magnetic field strength over time, and relativistic effects have to be taken into account. These various effects result in a different measured cyclotron frequency from the true $\nu_{c}$ given by equation~\eqref{eq:1}. The resulting frequency shift modifies the measured frequency ratio $R_{meas.}$:
\begin{equation}\label{eq:8}
R_{meas.} = \frac{\nu_{c,cal}+\Delta \nu_{c,cal}}{\nu_{c}+\Delta \nu_{c}}
\end{equation} 
from the ideal frequency ratio \(R_{ideal} = \nu_{c,cal}/\nu_{c}\).

The large value of the cyclotron frequency, in the MHz range, compared to the frequency shifts $\Delta \nu_{c}$, in the
Hz range, allows one to state that \(\Delta \nu_{c}/\nu_{c} \ll 1\). This consequently leads to a relative frequency ratio shift of
\begin{equation}\label{eq:9}
\frac{\Delta R}{R} = \frac{R_{meas.} - R_{ideal}}{R_{ideal}}  = \frac{\Delta \nu_{c,cal}}{\nu_{c,cal.}} - \frac{\Delta \nu_{c}}{\nu_{c}}. 
\end{equation} 
For most systematic effects studied in this paper, \(\Delta \nu_{c,1} \approx \Delta \nu_{c,2} = \Delta \nu_{c} \), therefore the relative frequency ratio shift will typically have the form \(\Delta R/R = (2 \pi \cdot \Delta \nu_{c}/B) \cdot \Delta(m/q) \), where we defined \(\Delta(m/q) := m_{cal.}/q_{cal.} - m/q\). From equation~\eqref{eq:9}, two main conclusions are drawn. Firstly, relative frequency ratio shift are in general smaller than the individual relative frequency shifts. Secondly, by measuring the frequency ratio of two species of similar mass-to-charge ratio $m/q$, one can reduce $\Delta R/R$. 

\section{The TITAN mass measurement Penning trap} \label{Sec:4}

\begin{figure}
  \begin{center}
    \includegraphics[width=0.80\textwidth]{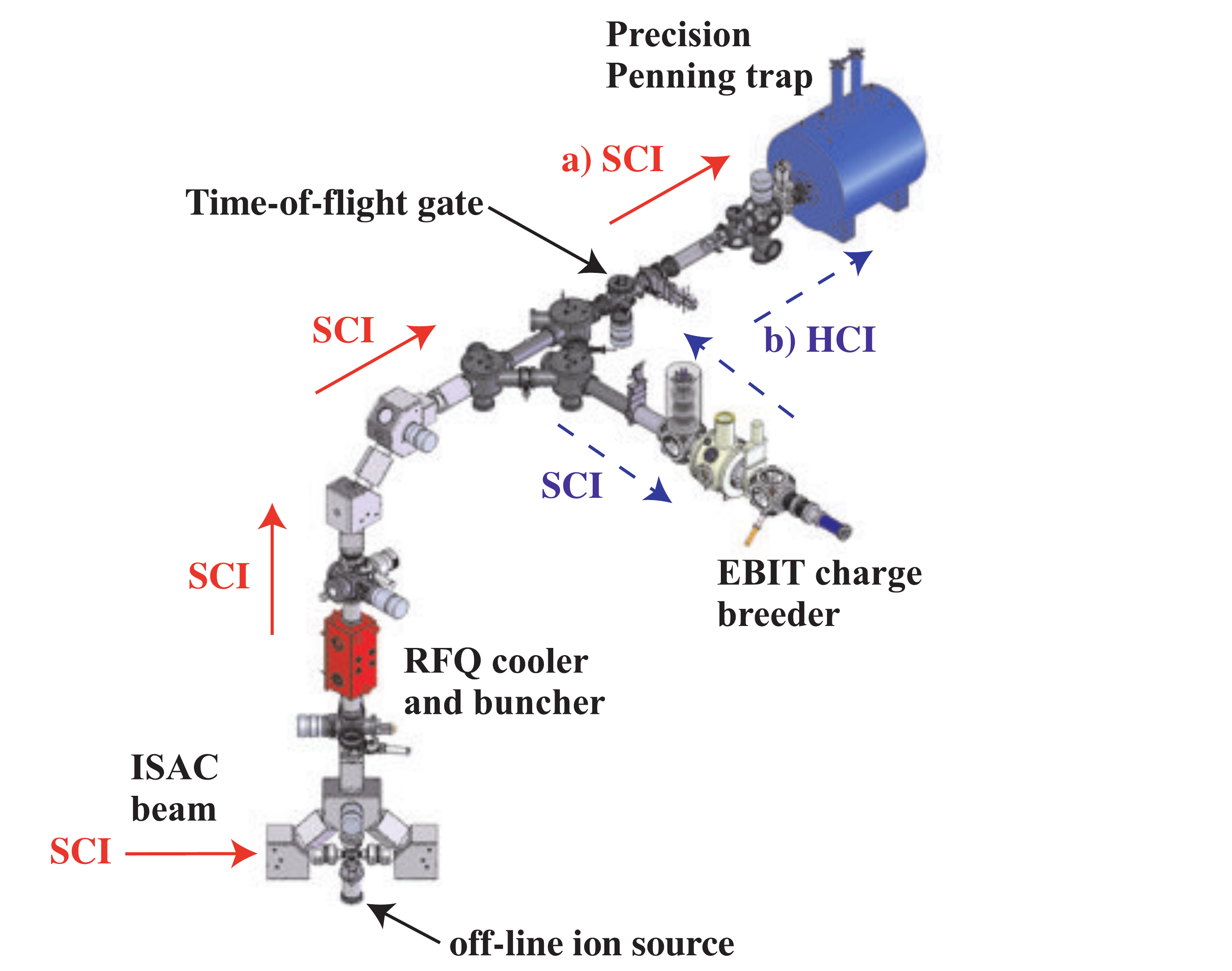}
    \caption{(colour on-line) The TITAN experimental setup which includes an RFQ, a high-precision Penning trap, an EBIT, a time-of-flight gate and an off-line ion source. a) Shown in solid-red is the path of the beam when mass measurement on singly charged ions (SCI) is performed. b) In dashed-blue is the path for mass measurements on highly charged ions (HCI).}{\label{fig:TITAN_setup}}
  \end{center}
\end{figure}
The high-precision mass measurements carried out at TITAN (shown in figure~\ref{fig:TITAN_setup}) are achieved through a series of steps. First, the continuous ion beam from ISAC (Isotope Separator and ACcelerator) \cite{Dom02} is delivered to TITAN where it is cooled and bunched using a gas-filled linear radio-frequency quadrupolar (RFQ) trap \cite{Smi06}. The subsequent step varies depending on whether a mass measurement is performed using singly charged ions (SCI), or highly charged ions. The choice depends on the required precision, half-life and production yield. The ions can either be transferred to an electron-beam ion trap (EBIT) (\cite{Sik05}-\cite{Lap10}), where charge breeding takes place (blue path in figure~\ref{fig:TITAN_setup}), or sent directly to the Penning trap (MPET) where the mass of the ion of interest is determined (red path in figure~\ref{fig:TITAN_setup}).

\begin{figure}
  \begin{center}
    \includegraphics[width=0.6\textwidth]{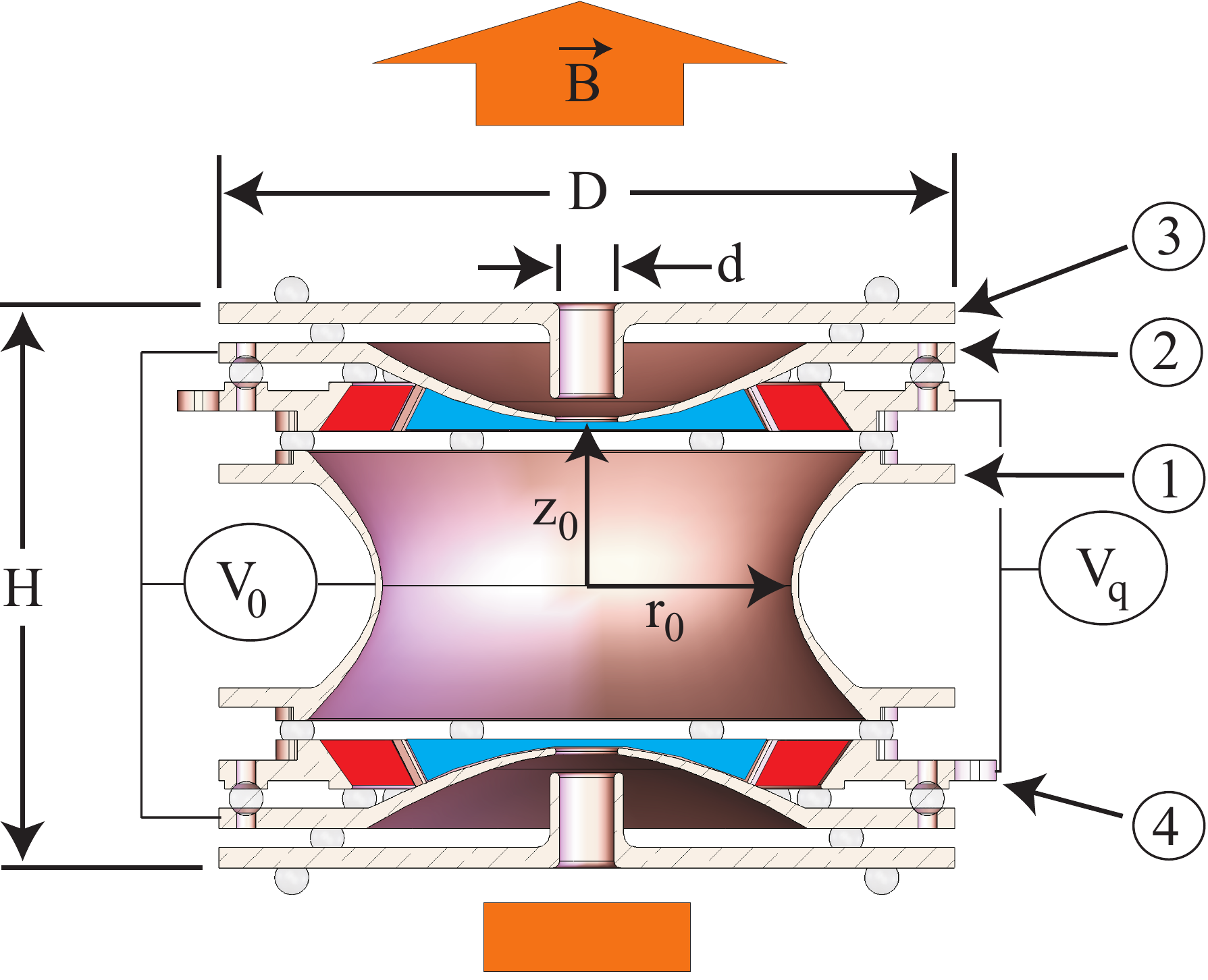}
    \caption{(colour on-line) Schematic of the TITAN Penning trap electrode configuration formed by the hyperbolic ring (labeled (1)), end cap electrodes (2), tube (3) and guard (4) correction electrodes. The RF is applied on (4) and the blue-red color code express the opposite phases of a quadrupolar excitation. The characteristic dimensions are given in table~\ref{tab:trapcharac}.}{\label{fig:MPETdimensions}}
  \end{center}
\end{figure}
\begin{table}[ht]
\begin{center}\caption{\label{tab:trapcharac} Characteristic dimensions of the TITAN Penning trap shown in figure~\ref{fig:MPETdimensions}.}
\begin{tabular}{|c|c|c|c|c|clcl}
\hline
Dimension & Value (mm) \\
\hline \hline
$r_{0}$ & 15 \\
\hline
$z_{0}$ & 11.785 \\
\hline
$d_{0}$ & 11.21 \\
\hline
$H$ & 41.48 \\
\hline
$D$ & 61 \\
\hline
$d$ & 4 \\
\hline
\end{tabular}
\end{center}
\end{table}
The TITAN Penning trap is shown in figure~\ref{fig:MPETdimensions}, with characteristic dimensions given in table~\ref{tab:trapcharac}. The trap is composed of two hyperboloids of revolution forming one ring (label (1) in figure~\ref{fig:MPETdimensions}) and two so-called end-cap electrodes (2). The ions are axially trapped by a harmonic quadrupole electrostatic potential produced by a potential difference, $V_{0}$, between the ring and the end cap electrodes, as shown in figure~\ref{fig:MPETdimensions}. Some anharmonicities in the trapping potential are introduced by the holes in the end-cap electrodes and by the finite size of the hyperbolic electrodes. Two sets of correction electrodes (labeled (3) and (4) in figure~\ref{fig:MPETdimensions}), are used to compensate for higher-order electric field components (for more detail see section~\ref{Sec:4c}). The radial confinement is provided by a magnetic field $B$. This section describes how various sources of systematic errors effects are studied and minimized. 

\subsection{Spatial magnetic field inhomogeneities} \label{Sec:4a}

For an ideal Penning trap, one assumes that the magnetic field strength is constant across the trapping region, i.e. \(B(x,y,z) = B_{0}\). For real traps, however, magnetic field inhomogeneities are created by the finite size of the solenoid and magnetic field distortion due to the magnetic susceptibilities of the trap material \cite{Bol90} or imperfections in the solenoid due to the finite size of the coil wire. In all cases, the lowest-order contribution to the magnetic field axial projection $B_{z}$ inhomogeneities has a quadrupole component, as seen in
\begin{equation}\label{eq:10}
B_{z}(z,r) = B_{0} \left \{ 1 + \beta_{2} \left ( z^{2} - r^{2}/2 \right ) \right \}
\end{equation}
where $z$ is the ion axial oscillation amplitude, $r$ is the ion radial position, $B_{0}$ is the unperturbed magnetic field strength and $\beta_{2}$ is the strength of the lowest order inhomogeneity component. This constant has been emperically determined for various other systems and typically ranges from 10$^{-10}$ to 10$^{-6}$ mm$^{-2}$ for TOF-ICR Penning traps \cite{Bol90} and \cite{Rin09b}.

The specifications for the construction of the TITAN magnet were such that the $B_{0}$ = 3.7 T field is homogenous within $\Delta B_{z}/B_{0}$ = 1 part-per-million (ppm) inside a 2 cm long by 1 cm diameter cylinder about the trap centre. Also, the trap electrode structure was manufactured by minimizing the required amount of material to realize the geometry. Moreover, only material of low magnetic susceptibility such as high conductivity oxygen-free copper for the electrodes and sapphire for the insulators was used. 

With these considerations, one can estimate the effect of the magnetic field inhomogeneities on the measured frequency ratio. The frequency shift from the magnetic field inhomogeneity \eqref{eq:10} is \cite{Bol90}:
\begin{equation}\label{eq:11}
\Delta \nu_{c} = \beta_{2} \nu_{c} \left \{ \left ( z^{2} - r_{+}^{2} \right ) - \frac{\nu_{-}}{\nu_{c}} \left ( r^{2}_{+} + r^{2}_{-} \right ) \right \},
\end{equation}  
where $r_{+}$ and $r_{-}$ are respectively the ion reduced cyclotron and magnetron radii.
Because  \( \nu_{c} \gg  \nu_{-} \), the term in braces is weakly mass dependent, leading to an overall effect on the cyclotron frequency ratio of:
\begin{equation}\label{eq:12}
\left ( \frac{\Delta R}{R} \right )_{mag.inhom.} \simeq -\beta_{2} \left ( r^{2}_{+} + r^{2}_{-} \right ) \frac{V_{0}}{2 B^{2} d_{0}^{2}} \Delta (m/q),
\end{equation}  
where we used \(\nu_{-} \approx V_{0}/(4 \pi B d_{0}^{2} )\).
The size of the TITAN Penning trap end cap hole can be used as estimate of the maximum workable $r^{2}_{+}$ or $r^{2}_{-}$, from which we estimate $\left ( r^{2}_{+} + r^{2}_{-} \right )$ $<$ 4 mm$^{2}$.  Using this estimate and assuming $\beta_{2}$ =  1 $\times$ 10$^{-6}$ mm$^{-2}$, one gets an absolute value in the shift in the frequency ratio of  $(\Delta R/R)_{mag.inhom.}$ $<$ 1.2 $\times$ 10$^{-11}$ V$^{-1}$ $\cdot V_{0} \cdot \Delta (m/q)$, where $\Delta (m/q)$ is in  units of amu/charge. Assuming $V_{0}$ = 35.7 V, the uncertainty due to the spatial magnetic field inhomogeneities becomes $(\Delta R/R)_{mag.inhom.}$ $<$ 4.3 $\times$ 10$^{-10} \cdot \Delta (m/q)$, which is one order of magnitude below the standard precision aimed for the TITAN Penning trap system. 

\subsection{Harmonic distortion and misalignment of the magnetic field axis} \label{Sec:4b}

The ideal Penning trap assumes a perfect alignment between the trap electrode structure axis and magnetic field axis (i.e. \(\vec B = B_{0} \hat{z}\)). It also assumes that the electrodes are aligned with respect to each another and without surface imperfections.
\begin{figure}[ht]
  \begin{center}{
    \includegraphics[width=0.8\textwidth]{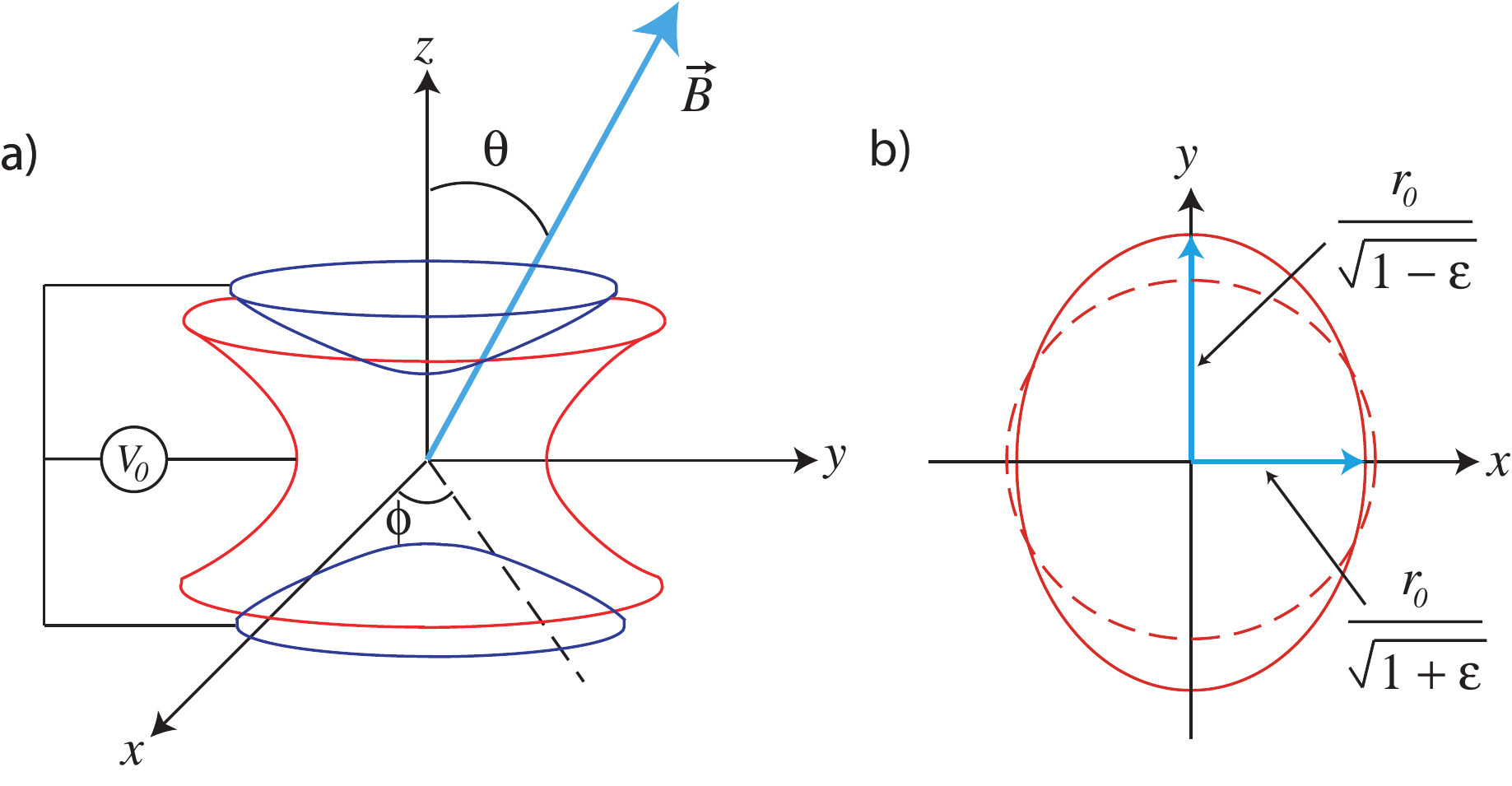}}
    \caption{a) Schematic of the
    electrode structure alignment with the magnetic field axis. b) Top view of a elliptically distorted ring electrode that leads to a non-zero asymmetry parameter $\epsilon$. Also shown is the undistorted ring (dashed lines).}
    {\label{fig:misalign}}
  \end{center}
\end{figure}
In reality (figure~\ref{fig:misalign} (a)), the magnetic field could have some misalignment with the trap axis.
Also, the trap electrodes could have deformations as shown in figure~\ref{fig:misalign} (b), and be misaligned with respect to one another. This would lead to a finite asymmetry parameter $\epsilon$, resulting in a distorted potential  \cite{Bro82} given by 
\begin{equation}\label{eq:13}
V_{harm.dist.} = \frac{V_{0}}{4 d_{0}^{2}} \left \{ ( 1 + \epsilon ) x^{2} + ( 1 - \epsilon ) y^{2}  \right \}. 
\end{equation}
These two imperfections modify the equation of motion of the ion in the trap resulting in a change of their eigen frequencies, modifying the measured cyclotron frequency according to \cite{Bro82}:
\begin{equation}\label{eq:14}
\Delta \nu_{c} =  \left ( \frac{9}{4} \theta^{2} - \frac{1}{2} \epsilon^{2} \right ) \cdot \overline{\nu}_{-},
\end{equation} 
where $\Delta \nu_{c}$ is the cyclotron frequency shift, $\overline{\nu}_{-}$ is the measured magnetron frequency and $\theta$ is the angle between the trap and magnetic field axis. The corresponding frequency ratio shift is given by:
\begin{equation}\label{eq:15}
(\Delta R/R)_{mis.} = \left ( \frac{9}{4} \theta^{2} - \frac{1}{2} \epsilon^{2} \right ) \cdot \left ( \frac{\Delta (m/q)}{m_{cal.}/q_{cal.}} \right ) \cdot \left ( \frac{\overline{\nu}_{-}}{\overline{\nu}_{+,cal}} \right ),
\end{equation}  
where $\overline{\nu}_{+,cal}$ is the measured reduced cyclotron frequency of the calibrant. By approximating \(\overline{\nu}_{+,cal}  \approx \overline{\nu}_{c,cal}\), equation~\eqref{eq:15} can be simplified to
\begin{equation}\label{eq:15b}
(\Delta R/R)_{mis.} = \left ( \frac{9}{4} \theta^{2} - \frac{1}{2} \epsilon^{2} \right ) \cdot \frac{V_{0}}{2 B^{2} d_{0}^{2}} \cdot \Delta (m/q).
\end{equation}  
The angle of misalignment $\theta$ was minimized by a precise alignment of the Penning trap electrode structure with the magnetic field axis using an electron beam. In addition, the system was built by requiring tight machining tolerances of 10$\mu$m for the trap electrodes and insulators. 
\begin{figure}
  \begin{center}
    \includegraphics[width=0.8\textwidth]{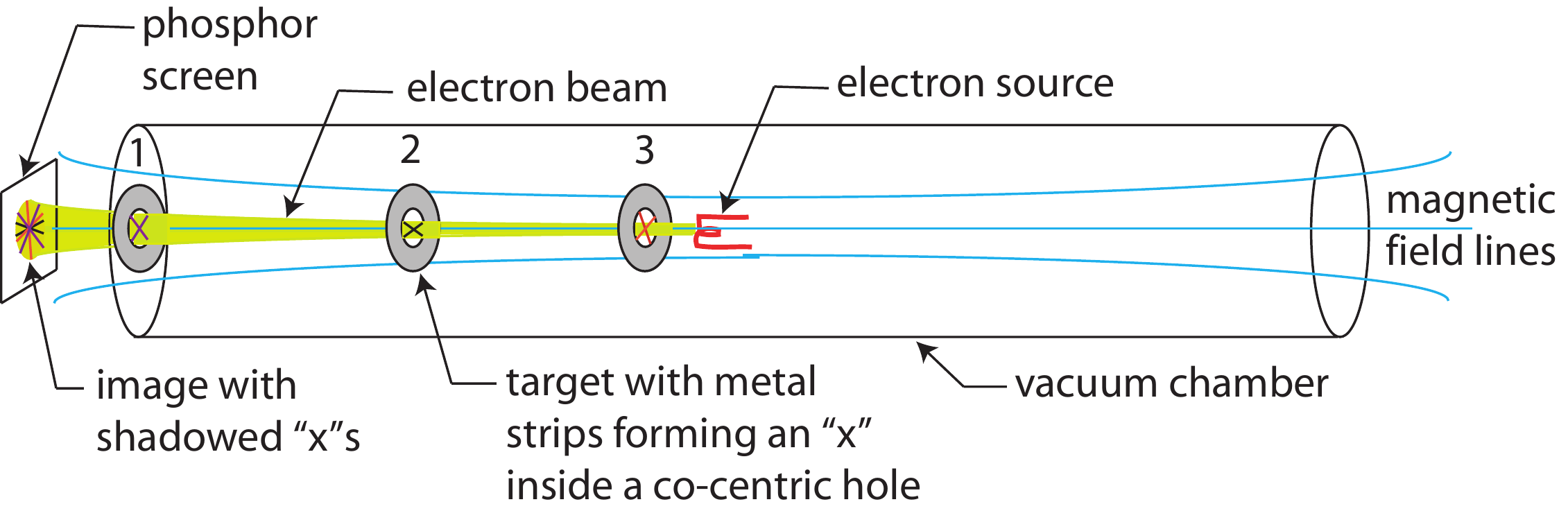}
    \caption{Schematic of the Penning trap vacuum
    chamber alignment using an electron source.}{\label{fig:Alignment}}
  \end{center}
\end{figure}
The trap vacuum chamber has been carefully aligned using an electron source positioned at the trap centre location, a phosphor screen placed at the end of the vacuum chamber, and three concentric targets as shown in figure~\ref{fig:Alignment}. The targets were made of an aperture with two metal stripes accurately positioned to form a cross within 0.01 mm of the hole centre. The stripes were 0.1 mm thick and the hole was 8 mm in diameter. The three targets were then secured in place along the Penning trap optics support frame, which in turn is centred along the vacuum tube ensuring the co-centricity of the targets. The vacuum tube is made using a pulled honed titanium tube with a very low tolerance on centricity of less than 0.13 mm over the length of 1.23 m. The tube acts as an optical bench for the trap structure. The cross patterns were rotated from one target to another as shown in figure~\ref{fig:Alignment} in order to facilitate the alignment of the chamber.

The alignment principle is based upon the fact that in the magnetic field, the electrons are guided along the field lines. A proper alignment of the vacuum chamber co-centric with the magnetic field lines is ensured when the three shadowed images of the targets metal strips are aligned and when a circular image of the electron beam spot is observed on the phosphor screen. Therefore, once the chamber has been adjusted such that the electron beam can pass through the three apertures, which are placed along the tube axis, a fine alignment is performed by aligning the ``shadows'' of the three crosses on a phosphor screen. This is done by moving the chamber with respect to the magnet housing in the x-y direction using a fine-thread external mechanical alignment mechanism.

\begin{figure}
  \begin{center}
    \includegraphics[width=0.5\textwidth]{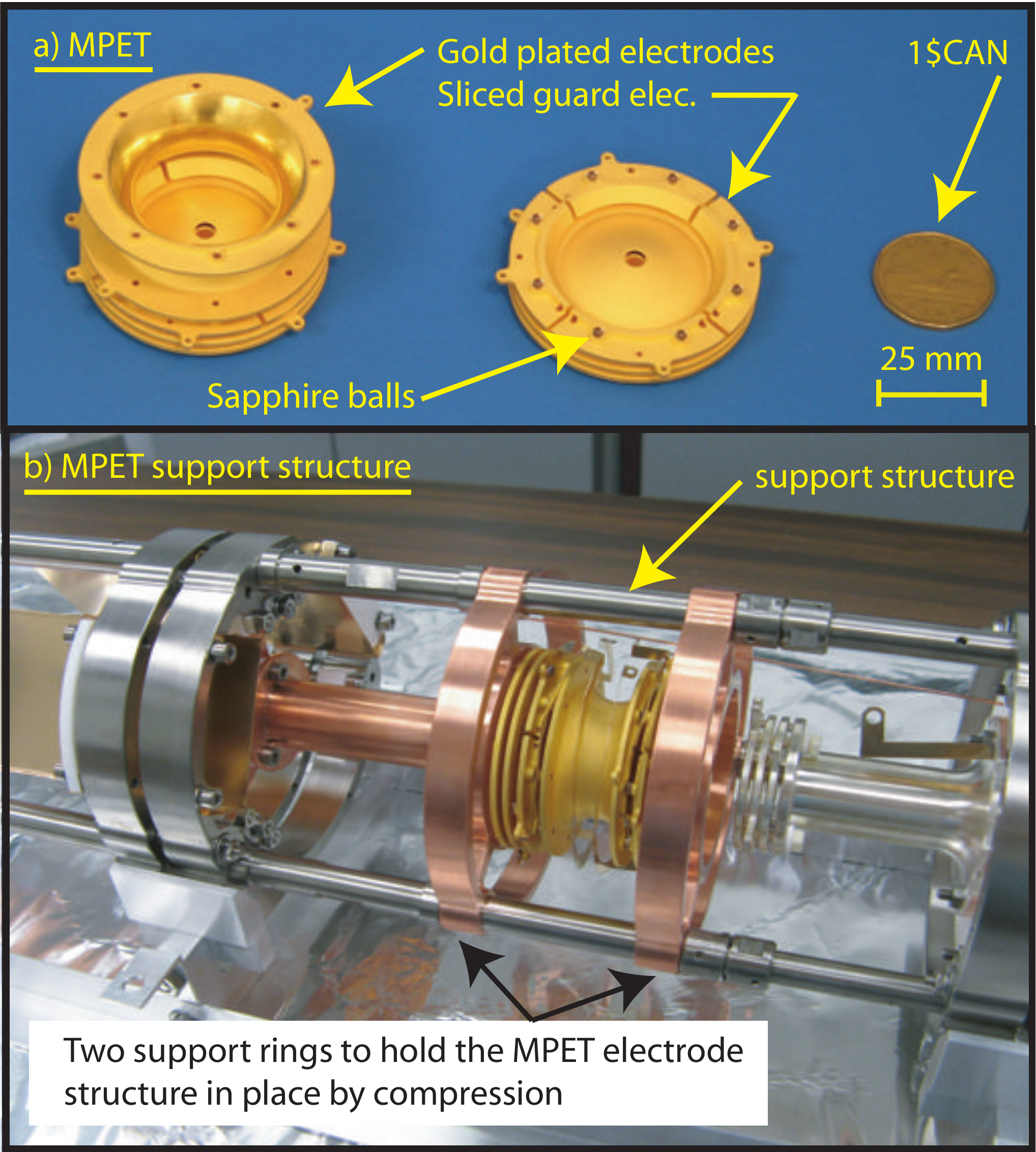}
    \caption{a) Left: one-piece ring electrode. Right: gold-plated Penning trap electrodes including the sapphire spheres (shown on top of the sliced guard electrode). Shown is a Canadian dollar coin for scaling. b) The TITAN mass measurement Penning trap (MPET) placed in the support structure frame. The trap structure is held in place by compression using two support rings.}{\label{fig:MPETphoto}}
  \end{center}
\end{figure}
The chamber has been finely adjusted until a displacement not worse than 0.04 mm between the three targets has been reached. Considering that the distance between aperture one and three (see figure~\ref{fig:Alignment}) is 590.5(1) mm, this gives an upper limit on the misalignment of the vacuum chamber with respect to the magnetic field axis of \(\theta_{chamber} < 7 \times 10^{-5}\) rad. 

\begin{figure}
  \begin{center}
    \includegraphics[width=0.7\textwidth]{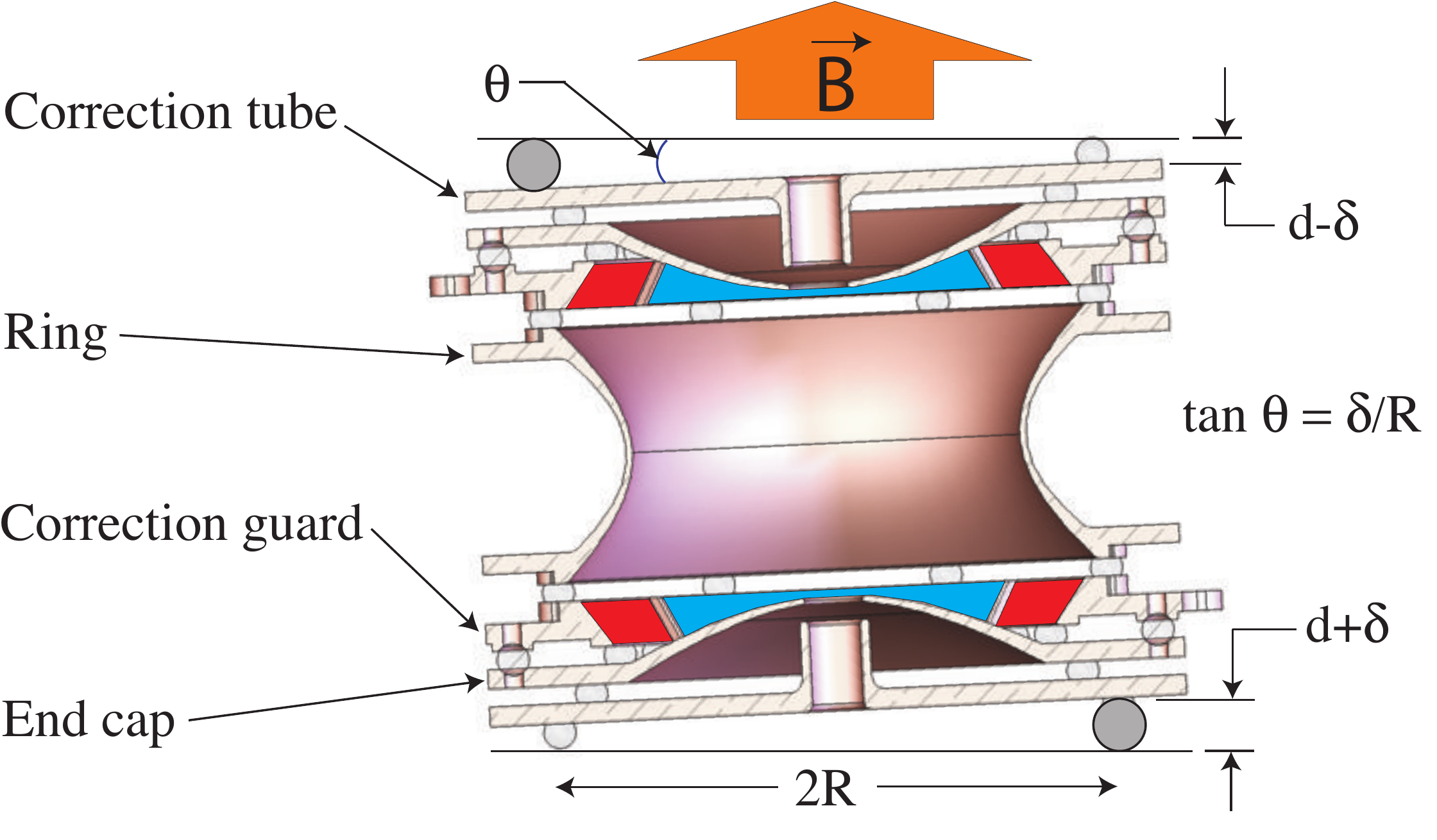}
    \caption{Schematic to demonstrate misalignment stemming from unequal sapphire sphere and holding hole size
    leading to a misalignment between the trap and magnetic field axes.}{\label{fig:SapphireSupport}}
  \end{center}
\end{figure}
An additional source of error comes from a misalignment of the trap electrode structure with respect to the support frame. The Penning trap electrodes are isolated from one another and from the support frame using sapphire spheres that are placed in a depression of the electrode. The entire structure is held in place by compression (see figure~\ref{fig:MPETphoto}), and the support structure is connected to the vacuum tube. The maximal misalignment of the trap with respect to the support structure happens when the sapphire balls are too wide, while the holes in which they are located are too small. This combination with extreme opposites for the mechanical tolerances on either side of the electrode structure, as shown in figure~\ref{fig:SapphireSupport}, would lead to a maximal shift. 

We consider sphericity tolerances on the sapphire balls of 5 $\mu$m and tolerances of 20 $\mu$m for the grove in which the balls are placed, which correspond to a doubling of the specified manufacturing tolerances. In the worst case, both the correction tube electrode with respect to the support frame and the end cap electrode with respect to the tube electrode are misaligned. Under these conditions, the misalignment is $\delta$ = 90 $\mu$m. Considering that the support sphere are placed along a circle of radius $R$ = 21.6 mm, the largest possible tilt is
\begin{equation}\label{eq:16}
\theta_{supp. max.} = \delta/R = 4.2 \times 10^{-3}.
\end{equation}  
Since the error stemming from the machining tolerances dominates (\(\theta_{supp.} \gg \theta_{chamber}\)) the error from the chamber alignment, the upper value on $\theta$, is $\theta_{max}$  = 4.2 $\times$ 10$^{-3}$. 
 
A non-zero asymmetry parameter, $\epsilon$, is caused by a number of effects. We discuss them and show how they
can be minimized. Firstly, localized oxidation patches on the surface of the electrodes cause undesired inhomogenous electric fields because of the different dielectric constant of the patch \cite{Tes97}. Such effects are minimized by gold-plating the trap electrode surfaces (see figure~\ref{fig:MPETphoto}). Secondly, the quadrupole deformation of the electric potential in the xy-plane is minimized by applying the RF-field on the correction guard electrodes to avoid splitting the ring electrode. Thirdly, misalignment of the ring electrode with respect to the trap axis is minimized by using high-tolerance sapphire spheres on which the trap electrodes sit and by requiring and ensuring tight electrode machining tolerance. 

Based on the manufacturing specifications, and the required machining tolerances, the maximum value for the ring electrode tilting angle $\alpha$ is found to be 0.0016 radians. Applying trigonometry and the equation for the ring electrode hyperbola, the asymmetry parameter $\epsilon$ due to tilted ring electrode is
\begin{equation}
\epsilon_{tilt} = \sin^{2} \alpha \cdot \left ( 1 + (r_{0}/z_{0})^{2} \right ) = 1.1 \times 10^{-5}.
\end{equation}

Lastly, elliptical deformation of the Penning trap electrodes would cause a non-zero asymmetry parameter $\epsilon$.
Elliptical deformation of the ring electrode will have the largest impact on the electrostatic potential, as it is the closest electrode to the trap centre. An elliptical deformation of the ring electrode (see figure~\ref{fig:misalign}(b)) corresponds to an elliptical Penning trap, a special case of hyperbolical Penning trap that has been studied extensively both theoretically \cite{Kre08} and experimentally \cite{Bre08}.  The elliptical deformation of the ring electrode can be described by: 
\begin{equation}\label{eq:harm6}
\frac{x^{2}}{(r_{0}/\sqrt{1+\epsilon})^{2}} + \frac{y^{2}}{(r_{0}/\sqrt{1-\epsilon})^{2}} = 1,
 \end{equation}
where $\epsilon$ gives an ellipticity that varies from 0 $<$ $\epsilon$ $<$ 1. Assuming the ring electrode radius from the trap centre is $r_{0}$ = 15 mm, and conservatively doubling the machining tolerance $\delta$ = 0.01 mm, we get   
\begin{equation}
\epsilon_{max} = \frac{4 \delta}{r_{0}} = 2.6 \times 10^{-3}.
\end{equation}
The error on the frequency ratio due to both the asymmetry parameter $\epsilon$ and the angle $\theta$ is given by equation~\eqref{eq:15b}. The maximal error on the frequency ratio is obtained when $\epsilon$ = 0 and $\theta$ = 4 $\times$ 10$^{-3}$ and is equal to \((\Delta R/R)_{mis.} = 1.2 \times 10^{-10} \mbox{V}^{-1} \cdot V_{0} \cdot \Delta (m/q) \). Using $V_{0}$ = 35.7V, one obtain an upper value of \((\Delta R/R)_{mis.} = 4.3 \times 10^{-9} \mbox{V}^{-1} \cdot \Delta (m/q) \).

\subsection{Non-harmonic imperfections of the trapping potential} \label{Sec:4c}

The holes in the end cap electrodes and the truncation of the Penning trap hyperboloid structure cause the trapping potential to be non-ideal and hence non-harmonic. Therefore one needs to consider octupole and dodecapole corrections to the trapping potential \cite{Bol90}. These corrections are given by:
\begin{eqnarray}\label{eq:21}
V_{4}(r,z) =  C_{4} \left ( \frac{V_{0}}{2 d_{0}^{4}} \right ) \left \{ z^{4} - 3 z^{2} r^{2} + \frac{3}{8} r^{4} \right \} \\\label{eq:real4b}
V_{6}(r,z) =  C_{6} \left ( \frac{V_{0}}{2 d_{0}^{6}} \right ) \left \{ z^{6} - \frac{15}{2} z^{4} r^{2} + \frac{45}{8} z^{2} r^{4} - \frac{5}{16} r^{6} \right \}, 
\end{eqnarray}   
where $C_{4}$ and $C_{6}$ are the octupole and dodecapole correction strengths, respectively. Because of the $1/d^{l}$ dependence of the potential, the contribution of higher-order terms become increasingly smaller. The procedure to calculate the frequency shifts due to the non-harmonicities is given in detail in \cite{Bro86} and \cite{Bol90}, and result in shifts in the radial eigenfrequencies of:
\begin{eqnarray}\label{eq:22}
\Delta \nu_{\pm} \approx  \pm \frac{3}{4} \frac{C_{4}}{d_{0}^{2}} \nu_{-} \left \{ ( r_{\pm}^{2} + 2 r_{\mp}^{2} ) - 2 z^{2} \right \} \\\label{eq:22b}
\Delta \nu_{\pm} \approx \pm \frac{15}{16} \frac{C_{6}}{d_{0}^{4}} \nu_{-} \left \{ -3 z^{4}+6z^{2} ( r_{\pm}^{2} + 2 r_{\mp}^{2} ) - (r_{\pm}^{4} + 3 r^{4}_{\mp} + 6 r_{+}^{2} r_{-}^{2}) \right \},
\end{eqnarray}
for the octupole and dodecapole terms, where the frequency shifts are nearly mass independent. This is because the magnetron frequency is very weakly mass-dependent \cite{Kon95}. Substituting \(\nu_{c} = \nu_{+} + \nu_{-}\) gives:
\begin{equation}\label{eq:23}
\Delta \nu_{c}  \approx  \frac{3}{4} \frac{( r_{-}^{2} - r_{+}^{2} )}{d_{0}^{2}} \nu_{-} \{ C_{4} +  \frac{5}{2} \frac{C_{6}}{d^{2}} \left ( 3 z^{2} - r_{+}^{2} - r_{-}^{2} \right ) \}.
\end{equation}

Assuming that no compensation voltage is applied on the correction-tube and -guard electrodes (see figure~\ref{fig:SapphireSupport}), one obtains the following coefficients: $C_{4}$ = 0.004 and $C_{6}$ = -0.082 for the TITAN Penning trap. Using a representative oscillation amplitude of $z$ = 3 mm and $\left ( r^{2}_{+} + r^{2}_{-} \right )$ = 4 mm$^{2}$, we get \(\Delta R/R_{pot.inhom.} = 2.8 \times 10^{-9} \cdot V_{0} \cdot \Delta (m/q)\). This is over 20 times larger than any other previously discussed frequency shift and for a typical trapping potential of $V_{0}$ = 35.7 V, this can lead to a frequency ratio shift of 1$\times$10$^{-7}$. Therefore, in order to perform accurate mass measurements at the level of $\delta m/m \approx$ 5$\times$ 10$^{-9}$, it is necessary to minimize the non-harmonic coefficients. 

\subsection{Additional sources of systematic errors} \label{Sec:4d}

Other sources of systematic errors include magnetic field fluctuations in time, ion-ion interaction and relativistic effects. These effects have been discussed for the specific case of the $^{6}$Li mass measurement (see \cite{Bro09, Bro10} for a detailed description of these effects). The systematic error on the linear interpolation of the calibrant cyclotron frequency due to time-dependant fluctuations of the magnetic field was found to be 0.04(11) ppb/h \cite{Bro09}. In most 
cases, the calibrations are spaced by less than one hour. Therefore, this represent a small systematic error.

Another source of systematic error would come from fluctuations of the trapping potential over time. This effect was investigated by monitoring at the potential applied to the specific electrodes at two occasion over a two months interval. The largest observed change in potential were for the ring electrode: $\Delta k_{ring}$ = 0.0022(1) and the injection-side correction tube electrode $\Delta k_{tube}$ = -0.0009(1). Such change in $k_{tube}$ from the optimal $k_{tube}$ = 1.53(2), would results in frequency shift between $\eta$ = 0.5 and 1.5 of 3.4(4) mHz for $^{6}$Li$^{+}$, which is much lower than the 80(50) mHz change observed in the previous section. Therefore, trapping potential fluctuations are considered to have a small relative effect.

\section{Compensation of the Penning trap electrical potential} \label{Sec:5}

From the previous sections, the largest possible source of error on the measured frequency ratio would come from the non-harmonic terms in the trapping potential. Therefore, these terms need to be minimized as they can induce a large shift in the cyclotron frequency. This is achieved using the correction guard and tube electrodes shown in figure~\ref{fig:MPETdimensions}. 


The optimal correction guard and tube potentials  with minimal non-harmonic coefficients $C_{N>2}$ have been estimated through the chi-square minimization of the difference between the potential produced by the trap electrodes and a quadratic target potential:
\begin{equation}\label{eq:27}
\chi^{2} = \sum{ \left \{ V_{ax}(z) - (z/z_{0})^{2} \right \}^{2}}.
\end{equation}
The effective axial potential $V_{ax}$ is a linear combination of the axial potentials produced by the individual electrodes:
\begin{equation}\label{eq:26}
V_{ax}(z) = k_{cap} V_{cap}(z) + k_{ring} V_{ring}(z) + k_{guard} V_{guard}(z) + k_{tube} V_{tube}(z);
\end{equation}  
where $V_{i}(z)$ is the axial potential produced when 1V is applied on the surface of a given electrode and $k_{i}$ are the scaling coefficient determined by the chi-square minimization. The potentials $V_{i}(z)$ corresponding to the TITAN Penning trap geometry were obtained using the Laplace equation solving capabilities of the ion-optics simulation software SIMION \cite{Dah00}.  
\begin{table}[ht]
\begin{center}\caption{\label{tab:calcPot} Calculated normalized potential $k_{i}$ needed to be applied on the ring, correction tube and guard electrodes in order to optimally compensate over a range of 8 mm from the trap centre.}
\begin{tabular}{|c|c|c|c|c|c|}
\hline
$k_{ring}$ & $k_{tube}$ & $k_{guard}$ & $C_{4}$ & $C_{6}$ \\
\hline \hline
-0.786 & 1.640 & 0.078 & -7$\times$10$^{-6}$ & 5$\times$10$^{-5}$ \\
\hline
\end{tabular}
\end{center}
\end{table}
Upon solving equation~\eqref{eq:27} with $k_{cap}$ = 1, one obtains the optimal potentials shown in table~\ref{tab:calcPot}. The size of the residual $C_{4}$ and $C_{6}$ coefficients for the optimal potential configuration is obtained by the least-square regression of
\begin{equation}\label{eq:31}
V(z) = \frac{V_{0}}{2} \left ( C_{0} + \frac{C_{2}}{d^{2}} z^{2} + \frac{C_{4}}{d^{4}} z^{4} + \frac{C_{6}}{d^{6}} z^{6} \right ).
\end{equation} 
The resulting coefficients shown in table~\ref{tab:calcPot} are a factor 1000 smaller than the values presented earlier, and would result in a significant reduction of the cyclotron frequency shift.


\subsection{Penning trap compensation using a dipole excitation} \label{Sec:5b}

The usual procedure \cite{Bec09} to compensate the electrostatic potential of a Penning trap consists of measuring the reduced cyclotron frequency $\nu_{+}$ of the ion in the trap for two values of the axial oscillation amplitudes $z$: one with $z \sim$ 0 and the other with $z >$ 0. The optimal compensation is the one that minimizes the difference between these two reduced cyclotron frequencies:
\begin{equation}\label{eq:34}
\delta \nu_{+} = \nu_{+}(z = z_{0}) - \nu_{+}(z \sim 0).
\end{equation}

Since the ions are trapped dynamically, the amplitude of the oscillations can be controlled by the closing time of the trap. Assuming the correct energy, the ions will have their minimal kinetic energy once they reach the trap centre and if at this point the trap is closed, the axial oscillation amplitude of the ions should be minimized. However, if the trap is closed at different times, earlier or later, it results in larger axial oscillation amplitudes.


\begin{figure}
\centering
\mbox{\subfigure{\includegraphics[width=0.5\textwidth]{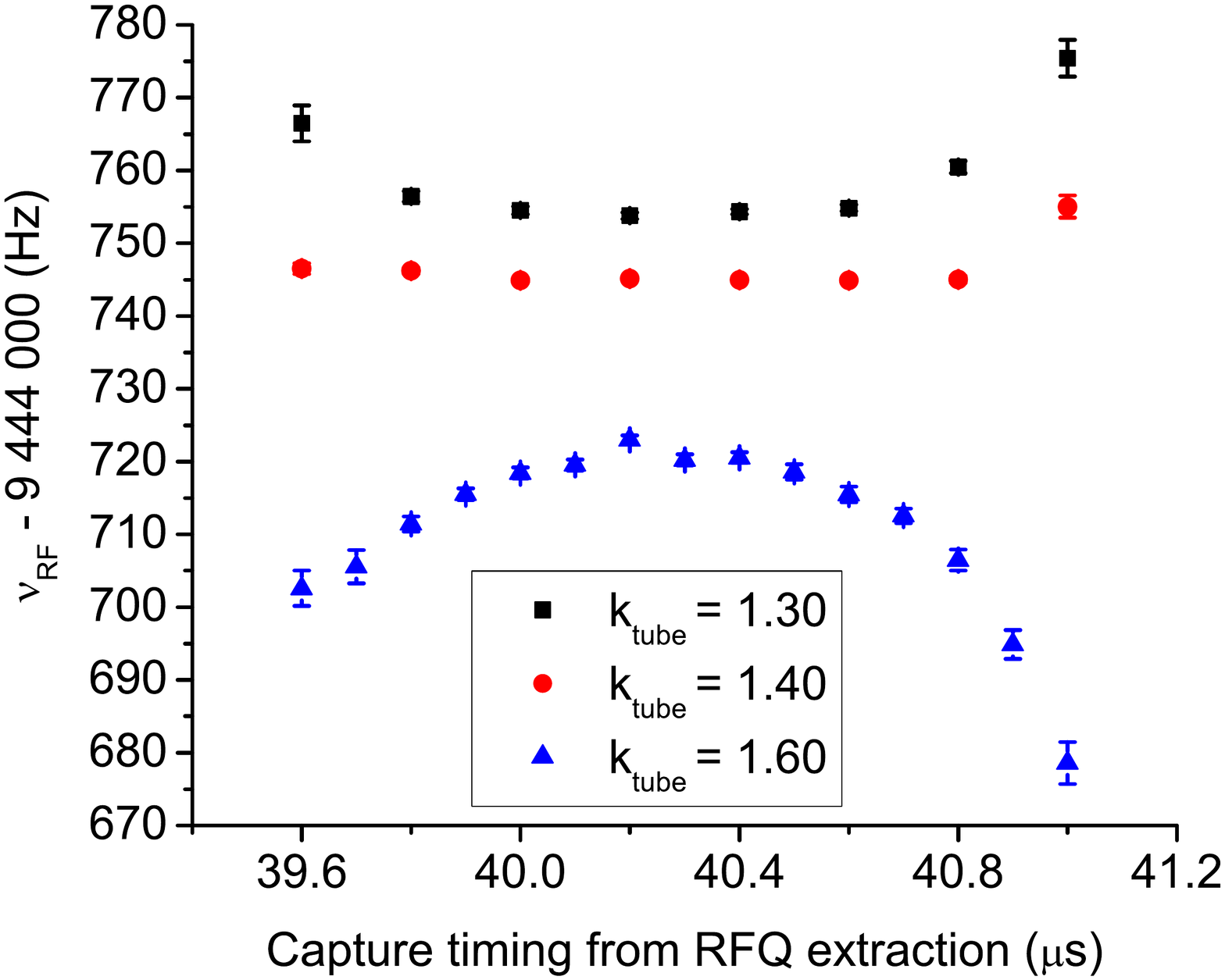}}\quad
    \subfigure{\includegraphics[width=0.5\textwidth]{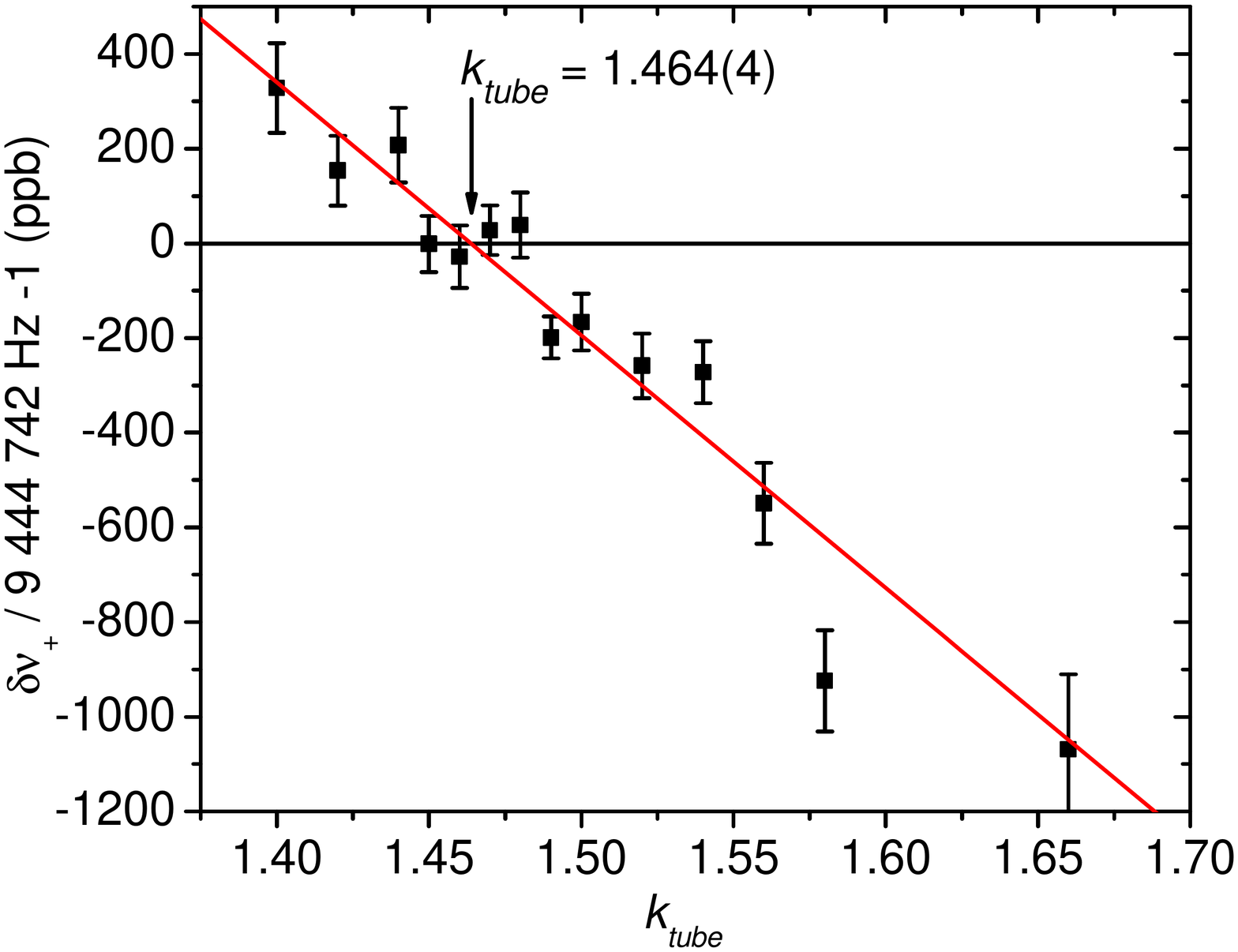}}}
    \caption{Left: Variation in the reduced cyclotron frequency as function of the capture time (hence position of the ions at the potential switching point) in the Penning trap for different correction tube settings $k_{tube}$ for $^{6}$Li$^{+}$. Note that close to $k_{tube}$ = 1.40, the variation in $\nu_{+}$ seems to be minimal. For these scans, $k_{guard}$ = -0.05. Right: Change in the reduced cyclotron frequency difference $\delta \nu_{+}$ of $^{6}$Li$^{+}$ with the correction tube potential $k_{tube}$ for a correction guard potential of $k_{guard}$ = -0.05. Note the linear change in $\delta \nu_{+}$ with $t_{cap}$. The linear fit crosses $\delta \nu_{+}$ = 0 for $k_{tube}$ = 1.464(4).}{\label{fig:TUBEfp}}
\end{figure}
From equations~\eqref{eq:22} and~\eqref{eq:22b}, the reduced cyclotron frequency changes in a quadratic form with an extremum at the trap centre. This is shown schematically in figure~\ref{fig:TUBEfp} (left) for different correction tube potentials. Note that the extremum in frequency found for capture time $t_{cap}$ = 40.3 $\mu$s (the capture timing is calculated from the extraction of the bunches from the RFQ) correspond to the time for which minimal ion axial oscillation amplitudes are observed. By changing either the correction guard or tube potential, one changes the values of the $C_{i}$ coefficients as expressed in equation~\eqref{eq:35}. This changes the amplitude and direction of the concavity of equations~\eqref{eq:22} and~\eqref{eq:23} (figure~\ref{fig:TUBEfp} (left)). 

Thus, the $C_{i>2}$ coefficients are minimized by changing the potential on the correction electrodes and taking the difference between the measured $\nu_{+}$ at the trap centre and the $\nu_{+}$ at a location away from the centre, i.e. \(\delta \nu_{+} = \nu_{+}(t_{cap} = 39.8 \mu \mbox{s}) - \nu_{+}(t_{cap} = 40.3 \mu \mbox{s}) \). 


Figure~\ref{fig:TUBEfp} (right) shows the linear behaviour in the reduced cyclotron frequency difference $\delta \nu_{+}$ with the correction tube voltage for $k_{guard}$ = -0.05. A linear regression of the data shows that $\delta \nu_{+}$ crosses zero for $k_{tube}$ = 1.464(4). This corresponds to one of the possible compensations. Therefore, by repeating the procedure for different correction guard potentials $k_{guard}$ one finds a family of different compensation settings. 

\subsection{Motivation for a compensation using two methods} \label{Sec:5a}


In this section we demonstrate that using only one electrostatic potential compensation method (such as the one presented in the previous section) leads to ambiguous values for the optimal correction tube and guard voltages, motivating the need for compensating the trapping potential using two different methods. 

In order to find the behaviour of the reduced cyclotron frequency difference $\delta \nu_{+}$ with the correction tube and guard potential, we investigated the $C_{4}$ and $C_{6}$ coefficients behaviour as function of $k_{tube}$ and $k_{guard}$. This was done by varying the scaling coefficients over the ranges -1.0 $<$ $k_{guard}$ $<$ 1.0 and 0.8 $<$ $k_{tube}$ $<$ 2.0 and by calculating the $C_{4}$ and $C_{6}$ coefficients using equation~\eqref{eq:31}.
\begin{figure}[ht]
  \begin{center}{
    \includegraphics[width=0.8\textwidth]{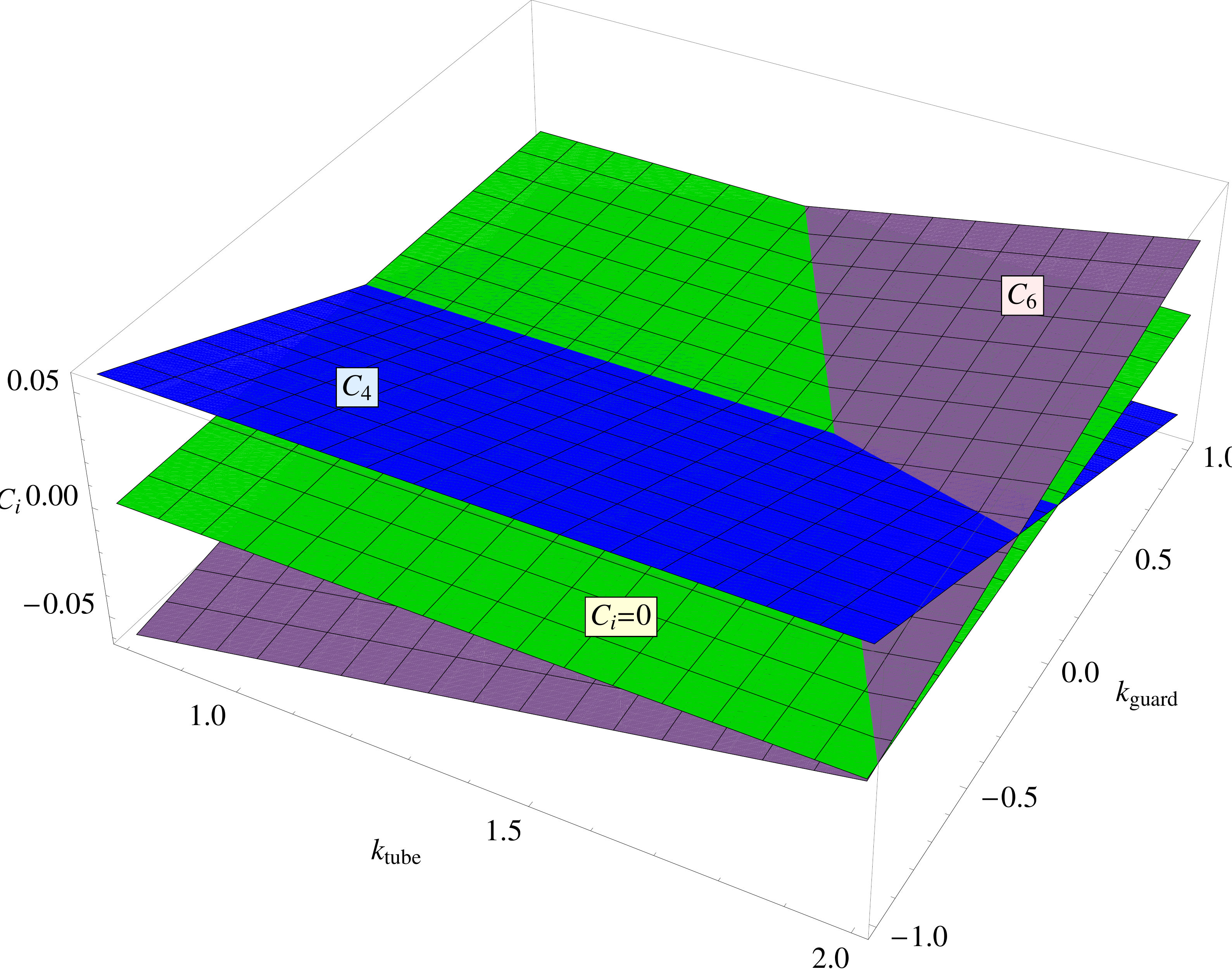}}
    \caption{Linear variation in the strength of the $C_{4}$ and $C_{6}$ coefficients as function of both 
    $k_{tube}$ and $k_{guard}$. Note that the planes cross at $k_{tube}$ = 1.64 and $k_{guard}$ = 0.08.}
    {\label{fig:C4C6behaviorGT}}
  \end{center}
\end{figure}
Figure~\ref{fig:C4C6behaviorGT} shows that the $C_{4}$ and $C_{6}$ coefficients vary linearly with both $k_{tube}$ and $k_{guard}$, giving rise to planar surfaces of equations
\begin{eqnarray}\label{eq:32}
C_{4} = 0.004 - 0.0003 k_{tube} - 0.051 k_{guard} \\\label{eq:33}
C_{6} = -0.083 + 0.050 k_{tube} + 0.017 k_{guard}
\end{eqnarray}  
in the $k_{tube}$-$k_{guard}$ space. As discussed in \cite{Bro86}, the octupolar term $C_{4}$ is mainly corrected by placing a correction guard electrode between the ring and end cap electrodes. This is confirmed by the strong dependance of $C_{4}$ with $k_{guard}$ shown in \eqref{eq:32}. Figure~\ref{fig:C4C6behaviorGT} shows that the dodecapole term $C_{6}$ is mainly affected by $k_{tube}$, confirming the literature \cite{Bol90}.

By inspecting equation~\eqref{eq:26} and equation~\eqref{eq:31}, it can be shown that the linear behaviour of $C_{4}$ and $C_{6}$ can also be generalized for higher order $C_{i}$:
\begin{equation}\label{eq:35}
C_{i} = a_{i} k_{tube} + b_{i} k_{guard} + c_{i} 
\end{equation}
which allows one to write the reduced cyclotron frequency difference as:
\begin{equation}\label{eq:36}
\delta \nu_{+} = \sum_{i=2}^{\infty}{a_{2i} h_{2i}} k_{tube} + \sum_{i=2}^{\infty}{b_{2i} h_{2i}} k_{guard} + \sum_{i=2}^{\infty}{c_{2i} h_{2i}}, 
\end{equation} 
where $h_{2i}$ are functions of the axial and radial positions of the ions in the trap that are not affected by variations of $C_{4}$ and $C_{6}$. The optimal compensation condition $\delta \nu_{+}$ = 0 leads to optimal values for $k_{tube}$ and $k_{guard}$, lying along a line given by the equation 
\begin{equation}\label{eq:38}
k_{tube} = -\frac{\sum_{i=2}^{\infty}{b_{2i} h_{2i}}}{\sum_{i=2}^{\infty}{a_{2i} h_{2i}}} k_{guard} - \frac{\sum_{i=2}^{\infty}{c_{2i} h_{2i}}}{\sum_{i=2}^{\infty}{a_{2i} h_{2i}}}.  
\end{equation} 
Since there can only be one sets of $k_{tube}$ and $k_{guard}$ that leads to a minimal value of the $C_{4}$ and $C_{6}$ coefficients (see figure~\ref{fig:C4C6behaviorGT}), and to an optimal compensation, one needs a second compensation approach that selects the correct setting along this line. This is achieved by carrying out two independant methods of compensating the trap. It should be noted that figure~\ref{fig:C4C6behaviorGT} does not include higher order terms which are present when $C_{4}$ and $C_{6}$ are effectively zero. However, the terms with the dominant contribution are minimized. 

\subsection{Penning trap compensation using a quadrupole excitation} \label{Sec:5c} 

The compensation using a quadrupole excitation consists of measuring the cyclotron frequency of the ion for two different conversion factors, $\eta$. This factor defines the sizes of the magnetron and reduced cyclotron radii ($r_{-}$ and $r_{+}$) at the end of the excitation phase and from equation~\eqref{eq:23}, changing these radii leads to different cyclotron frequencies. In this method the non-harmonic terms are minimized by finding the potential $k_{guard}$ and $k_{tube}$ that minimizes the change in the cyclotron frequency with $\eta$. The conversion factor itself is changed through a variation of the RF amplitude $V_{q}$ for constant excitation time $T_{q}$ (see equation~\eqref{eq:3b}).

The change of $\nu_{c}$ with $V_{q}$ was studied by numerically solving the equation of motion with an added $C_{4}$ term.
\begin{figure}[ht]
  \begin{center}{
    \includegraphics[width=0.7\textwidth]{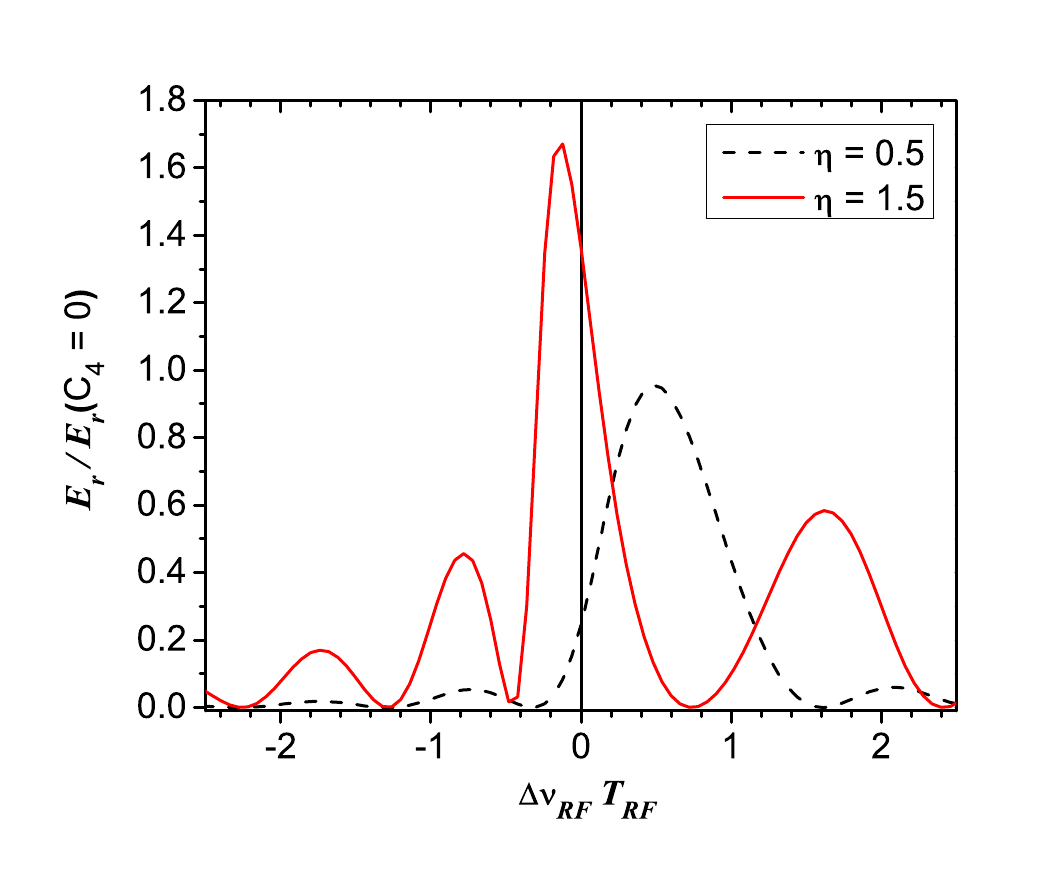}}
    \caption{Radial energy gain $E_{r}$ (calculated from equation~\eqref{eq:43}) as a function of the detuning $\Delta \nu_{RF} T_{RF}$ for conversion factor $\eta$ = 0.5 and 1.5 for $C_{4} T_{RF}$ = 0.01. Note that this is a large $C_{4}$ value used to demonstrate the effect. The energy gain was normalized to the radial energy obtained from $C_{4}$ = 0.}
    {\label{fig:ErvsC4}}
  \end{center}
\end{figure}
Note that the cyclotron frequency is also be modified by $C_{6}$ and higher order terms, but to simplify we only study the changes involving $C_{4}$. Upon solving these equations of motion, one obtains the radial energy profile as a function of the detuning frequency \(\Delta \nu_{q} = \nu_{q} - \nu_{c}\):
\begin{equation}\label{eq:43}
E_{r}(\Delta \nu_{q}) = \frac{1}{2} M \left ( \dot{x}(\Delta \nu_{q})^{2} + \dot{y}(\Delta \nu_{q})^{2} \right ),
\end{equation}  
where the dots denotes temporal derivatives and ($x$, $y$) is the position of the ion in the radial plane. Figure~\ref{fig:ErvsC4} shows that for $C_{4}$ $\neq$ 0, the radial energy profile is no longer symmetric. The deformation is more pronounced for over-converted resonances ($\eta$ $>$ 1), due to their smaller line width. For the under-converted case ($\eta$ $<$ 1), the centre frequency is more shifted. This is due to the ion magnetron motion being not fully converted into reduced cyclotron motion. Hence the ion then spent more time in regions where the $C_{i>2}$ components are larger, leading to a larger shift in the centroid frequency. 

\begin{figure}
\centering
\mbox{\subfigure{\includegraphics[width=0.5\textwidth]{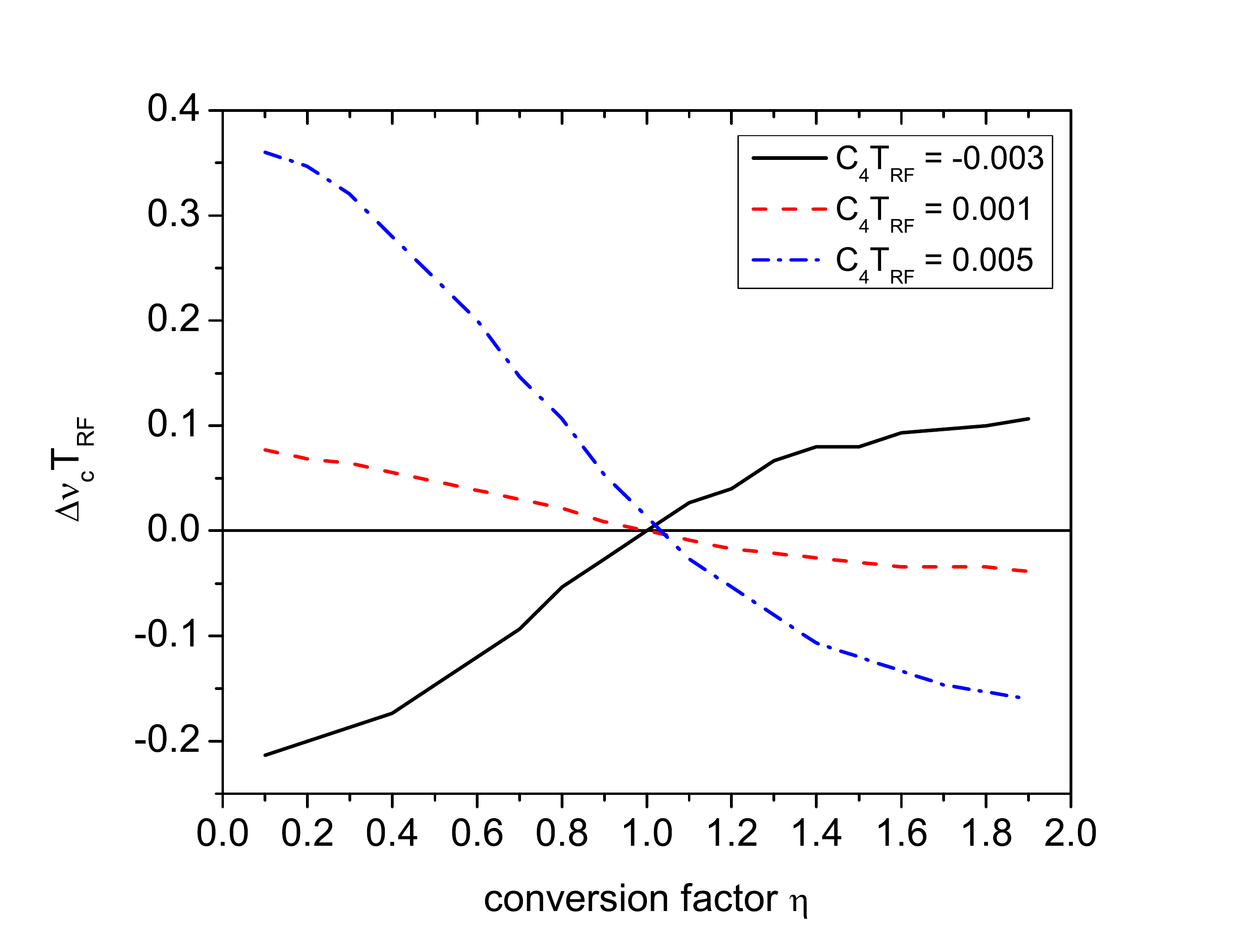}}\quad
    \subfigure{\includegraphics[width=0.5\textwidth]{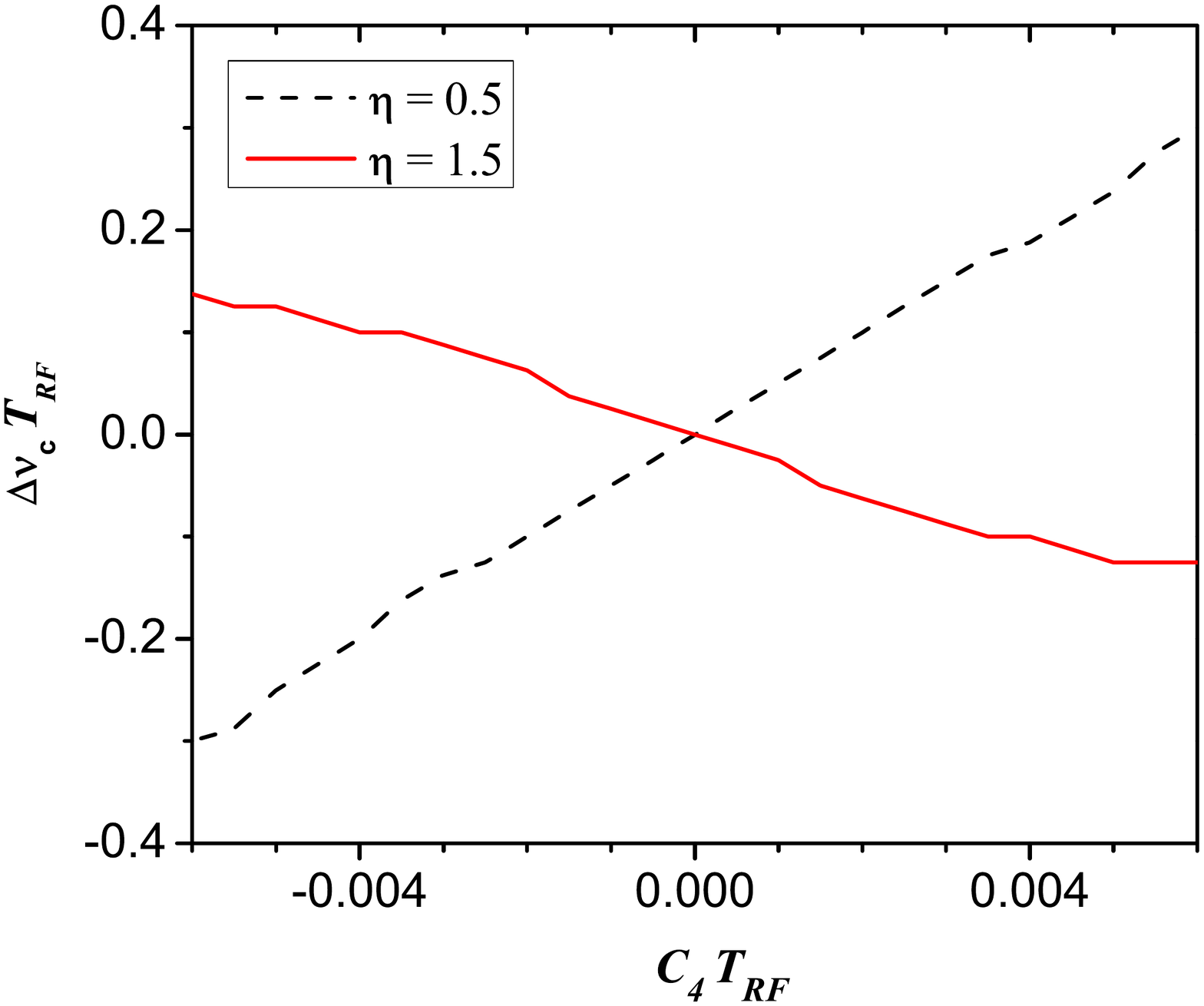}}}
    \caption{Left: Shift in the cyclotron frequency with the conversion factor $\eta$ for $C_{4} T_{RF}$ = -0.003, 0.001 and 0.005. Right: Change in the frequency shift $\Delta \nu_{c}$ as function of the octupole coefficient $C_{4}$
    with $\eta$ = 0.5, 1.5. These figures are numerical calculations results.}{\label{fig:fcshift}}
\end{figure}
Figure~\ref{fig:fcshift} (left) shows how the cyclotron frequency $\nu_{c}$ changes with $\eta$ for three different non-zero $C_{4}$ = -0.003, 0.001 and 0.005. As expected from equation~\eqref{eq:23}, the larger $C_{4}$, the more sensitive with $\eta$ the cyclotron frequency becomes. Also, the shift in frequency $\Delta \nu_{c}$ flips sign together with $C_{4}$. Figure~\ref{fig:fcshift} (right) gives a similar view of this phenomenon, except $C_{4}$ is varied for $\eta$ = 0.5 and $\eta$ = 1.5. This figure shows the only case where the cyclotron frequency is the same for the two conversion factor when $C_{4}$ = 0. It also shows that for values of $C_{4}$ close to zero, the cyclotron frequency change linearly with $C_{4}$ for both $\eta$ = 0.5 and $\eta$ = 1.5. Therefore, the optimal trap compensation that minimizes $C_{4}$ and other higher order terms will be equal to the correction guard $k_{guard}$ and tube $k_{tube}$ setting for which the cyclotron frequency difference for $\eta$ = 0.5 and $\eta$ = 1.5 equals zero.

\begin{figure}
\centering
\mbox{\subfigure{\includegraphics[width=0.5\textwidth]{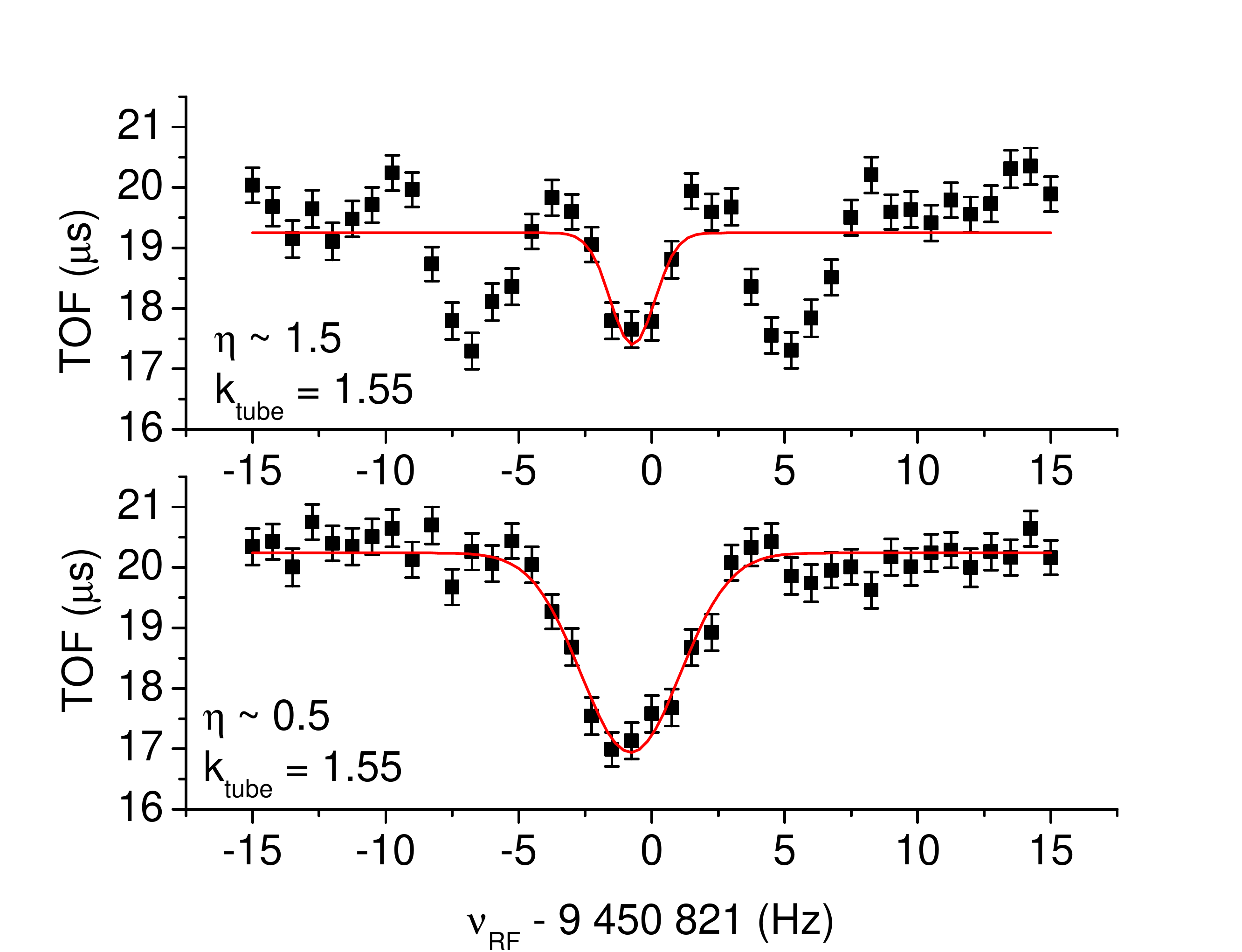}}\quad
    \subfigure{\includegraphics[width=0.5\textwidth]{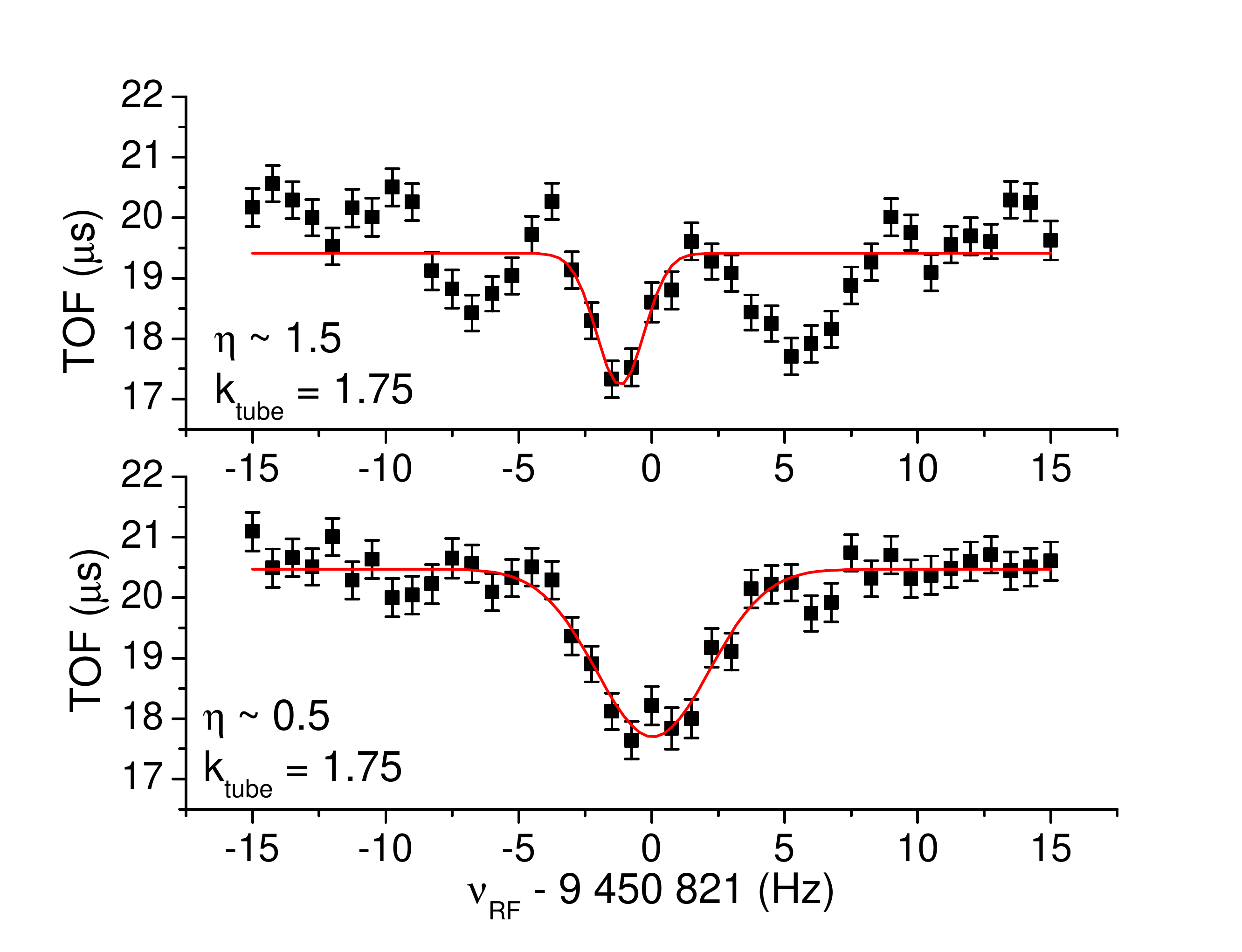}}}
    \caption{Over- ($\eta \sim$ 1.5) and under- ($\eta \sim$ 0.5) converted time of flight resonant spectra for  $k_{guard}$ = -0.05. Left: A close to optimal setting ($k_{tube}$ = 1.55) resulting in a symmetric over-converted TOF spectra. Right: A clearly non-optimal setting ($k_{tube}$ = 1.75) resulting in a crocked over-converted TOF spectra.}{\label{fig:TOFtube}}
\end{figure}
In the following, this compensation is performed by scanning the correction tube potential for three different correction guard voltages with values: $k_{guard}$ = -0.05, 0.06 and 0.08. For each $k_{tube}$, the cyclotron frequency was determined for $\eta$ = 0.5 and $\eta$ = 1.5 by fitting the centroid using a Gaussian curve as shown in figure~\ref{fig:TOFtube}. The previous numerical calculations results shown in figure~\ref{fig:ErvsC4}, reveal that in the presence of a large $C_{4}$ term in the electrostatic potential, the radial energy gain profile of the ion when $\eta \sim$ 1.5 becomes asymmetric about the cyclotron frequency. Figure~\ref{fig:TOFtube}~(right) shows that such effect gets translated into an asymmetric TOF resonance spectrum. When the compensation gets better, the TOF resonance spectrum when $\eta \sim$ 1.5 becomes more symmetric as shown in figure~\ref{fig:TOFtube}~(left). A non-optimal compensation also results in a large change in the cyclotron frequency for $\eta \sim$ 0.5, as shown in figure~\ref{fig:TOFtube}~(right). 

\begin{figure}[ht]
\centering
\mbox{\subfigure{\includegraphics[width=0.5\textwidth]{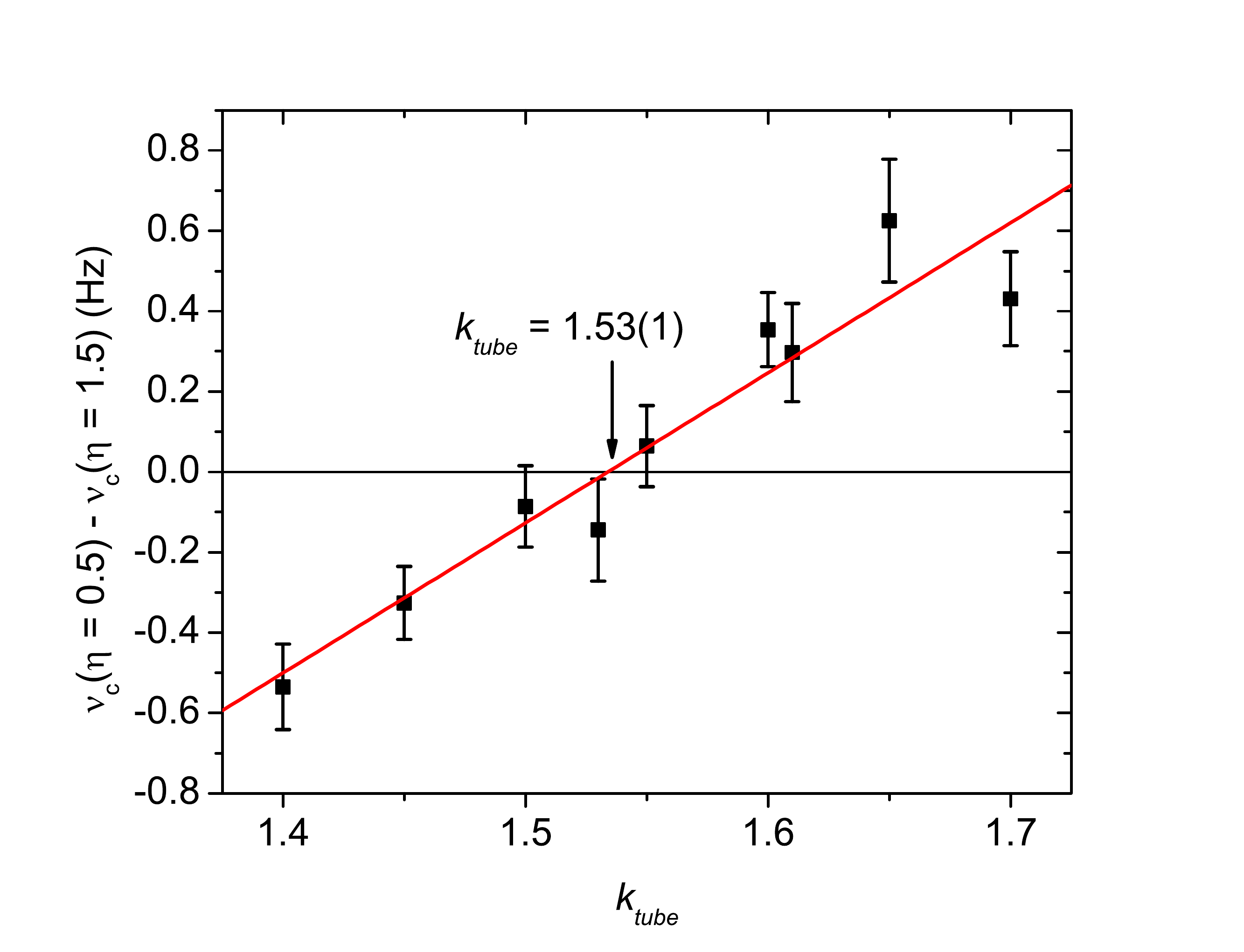}}\quad
    \subfigure{\includegraphics[width=0.5\textwidth]{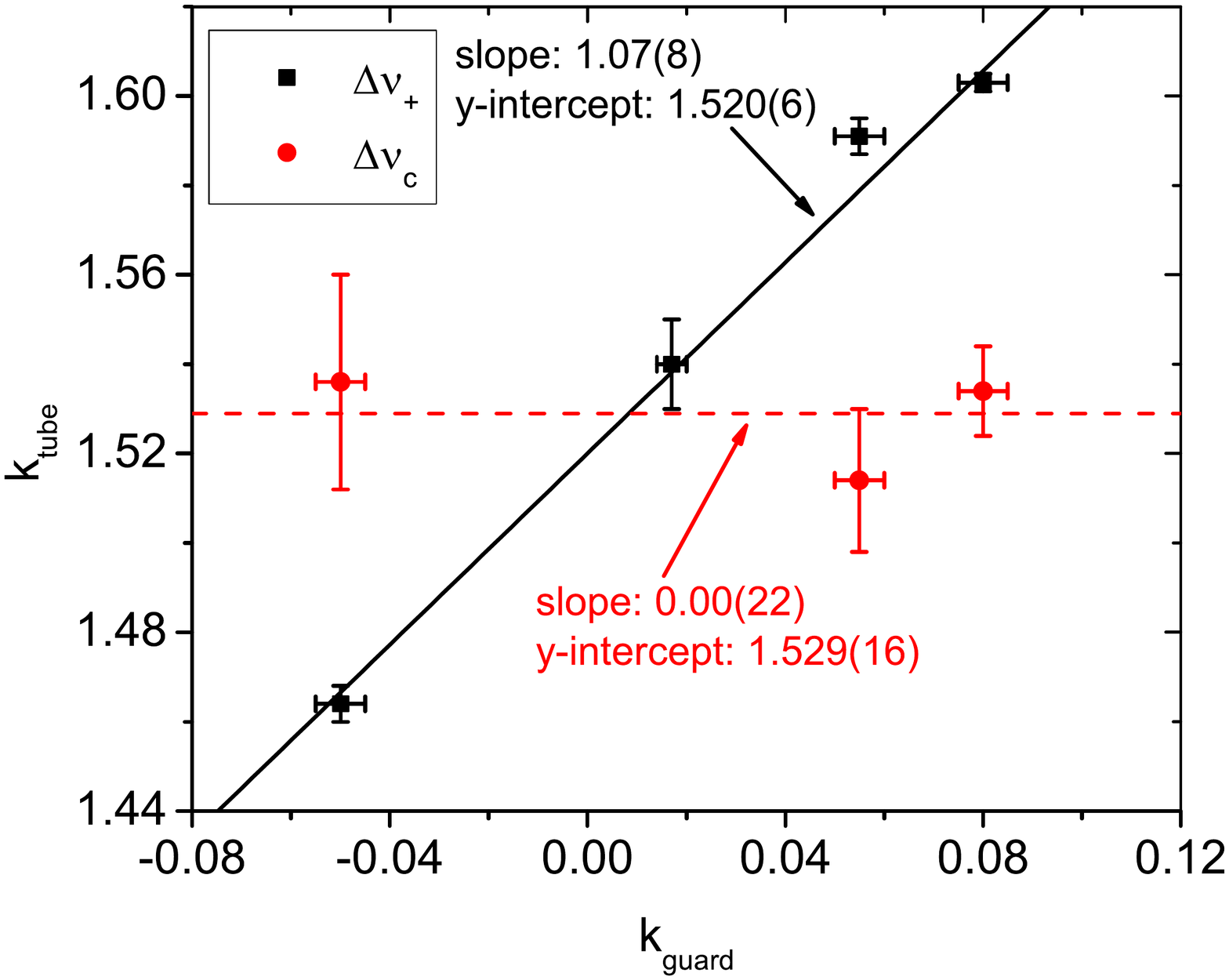}}}
    \caption{Left: Change in the fitted cyclotron frequency difference \(\nu_{c}(\eta = 0.5) - \nu_{c}(\eta = 1.5)\) with the scaled correction tube potential $k_{tube}$ for $k_{guard}$ = 0.08. Right: Optimal $k_{guard}$ and $k_{tube}$ found using a minimization of $\delta \nu_{+}$ and $\delta \nu_{c}$. The intersection of the solid and dashed lines correspond to the optimal combination of $k_{guard}$ and $k_{tube}$.}
    {\label{fig:TubeVrf}}
\end{figure}
Next we calculated the cyclotron frequency difference \(\delta \nu_{c} = \nu_{c}(\eta = 0.5) - \nu_{c}(\eta = 1.5)\), from which $\delta \nu_{c}$ = 0 was found from linear regression. The difference between the under-converted cyclotron frequency and the optimal over-converted cyclotron frequency for different correction tube voltage and using $k_{guard}$ = 0.06 is shown in figure~\ref{fig:TubeVrf}. The linear regression of $\delta \nu_{c}$ for this $k_{guard}$ yielded an optimal correction tube voltage of $k_{tube}$ = 1.53(1). Note that the two time-of-flight profiles at $k_{tube}$ = 1.52 and 1.54 for $\eta$ = 1.5 were the most symmetric, which is a clear indication of reaching the optimal compensation. 
     
The same procedure was repeated for correction guard voltages of $k_{guard}$ = 0.08 and -0.05.
The three different optimal $k_{guard}$ and $k_{tube}$ found using this method, together with a linear regression, are presented in figure~\ref{fig:TubeVrf} (right). This graph also shows the results from the compensation using the dipole resonances (see section 4.1). The optimal compensation of the non-harmonic terms in the trapping potential was taken as the intersect of the two lines. These two lines meet for $k_{tube}$ = 1.53(2) and  $k_{guard}$ = 0.01(2). For the typical end cap potential $V_{cap}$ = 20V used at TITAN, these corresponds to $V_{tube}$ = 30.6(4)V and $V_{guard}$ = 0.2(4)V.

\begin{figure}[ht]
  \begin{center}{
    \includegraphics[width=0.7\textwidth]{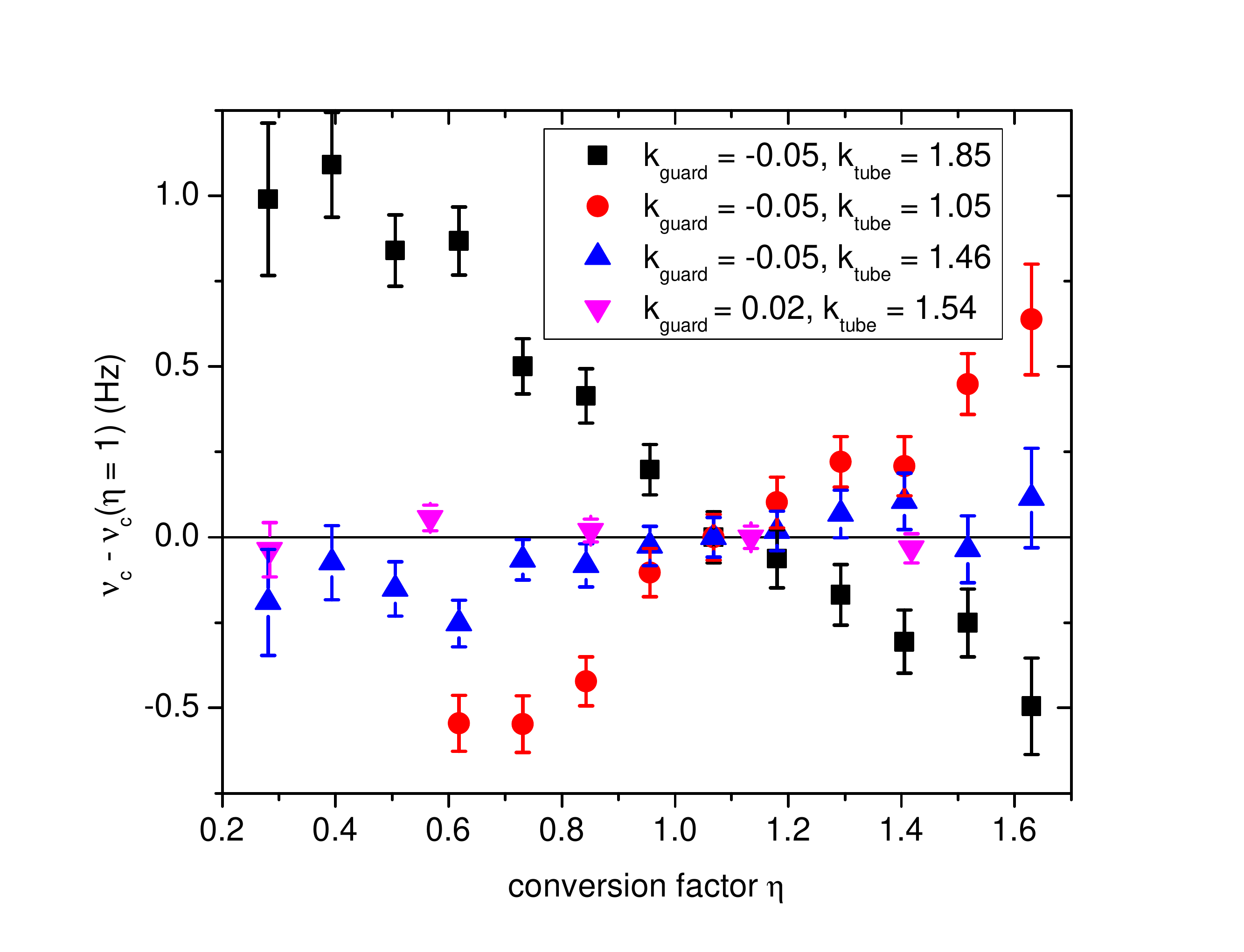}}
    \caption[$\nu_{c}$ as a function of $\eta$ for different correction potential.]
        {Fitted cyclotron frequency as a function of the conversion factor $\eta$ for different correction
    tube ($k_{tube}$) and guard ($k_{guard}$) potentials.}
    {\label{fig:VrfscanTG}}
  \end{center}
\end{figure}
The validity of this result was verified by investigating potential changes in the cyclotron frequency with the conversion
factor $\eta$. Figure~\ref{fig:VrfscanTG} shows that the changes in the cyclotron frequency is minimal for
$k_{tube}$ = 1.54 and  $k_{guard}$ = 0.02 compensation where the largest change in frequency with the conversion factor is 80(50) mHz. For comparison purposes, three other settings are shown. When $k_{tube}$ is very different from the optimal value, it induces a strong shift in the cyclotron frequency as the conversion factor is varied from about 0.3 to 1.6. As expected from figure~\ref{fig:fcshift}, this shift change direction for $k_{tube}$ values above or below the optimal one.

Figure~\ref{fig:VrfscanTG} also shows the setting $k_{tube}$ = 1.46 and  $k_{guard}$ = -0.05 which was previously found as being optimal using the minimization of $\delta \nu_{+}$ method. However, when varying the conversion factor $\eta$, changes in the cyclotron frequency of 260(80) mHz are observed, which are three times larger than the $k_{tube}$ = 1.54 and  $k_{guard}$ = 0.02 compensation. This observation confirms the relevance of compensating the non-harmonic terms of the potential by using two different observables. 

\section{Experimental determination of the mass-dependent frequency ratio shift} \label{Sec:5e}

The various systematic effects studied and minimized in section 3.1 to 3.3 all results in relative changes in the frequency ratio of the form \(\Delta R/R = (2 \pi \cdot \Delta \nu_{c}/B) \cdot \Delta(m/q)\). At the end of these sections, we presented upper limit estimates of $\Delta R/R$ due to the magnetic field inhomogeneities, the misalignment of the trap electrodes with the magnetic field, the harmonic distortion and the non-harmonic terms in the trapping potential. These estimates were based on the chosen trap geometry, machining tolerances and the trap alignment with the magnetic field. 

A more realistic value for the total contribution of the systematic effects for which the relative changes in the frequency ratio depends on $\Delta(m/q)$ can be evaluated experimentally by measuring the frequency ratio of ions with different mass-to-charge ratio. Using these frequency ratio, the atomic mass of one of the two species is calculated using equation~\eqref{eq:6} and compared to the most precise value from literature. The difference in mass \(\Delta m = m(\rm TITAN) - m(\rm literature)\) is then used to compute a combined systematic shift:
\begin{equation}\label{eq:44}
\frac{\Delta R}{R} \cdot \frac{1}{\Delta (m/q)} = \frac{\Delta m}{m} \cdot \frac{1}{\Delta (m/q)}.
\end{equation} 
\begin{table}[ht]
\begin{center}\caption{\label{tab:MassDepShift} Frequency ratios \(R = \nu_{c,cal}/\nu_{c}\) of $^{6}$Li \cite{Bro10}, $^{23}$Na, $^{39}$K, and $^{41}$K using as calibrant $^{7}$Li, H$_{3}$O, $^{23}$Na, and H$_{3}$O, respectively. Difference in the mass excess $\Delta m$ of $^{6}$Li, $^{23}$Na, $^{39}$K and $^{41}$K as measured with the TITAN and from the FSU \cite{Mou10} Penning traps. From these measurements the systematic error due to the incomplete compensation $\Delta \nu/(\nu A)$ is derived. $N$ is the number of measurements taken.}
\begin{tabular}{|c|c|c|c|c|c|}
\hline
Specie & $R \times$10$^{6}$ & $\Delta(m/q)$ & $\Delta m$ (eV) & syst. error (ppb/u) & $N$ \\
\hline \hline
$^{6}$Li & 857 332.053 6(37) & 1 & 1(24) & -0.1(2.2) & 43 \\
\hline
$^{23}$Na & 1 208 823.886(4) & 4 & -66(77) & -0.8(9) & 12 \\
\hline
$^{39}$K & 1 694 844 656(12) & 16 & -52(263) & -0.1(5) & 13 \\
\hline
$^{41}$K & 2 153 834.195(13) & 22 & -109(226) & -0.1(3) & 15 \\
\hline \hline
Total & & & & -0.2(2) & 40 \\
\hline
\end{tabular}
\end{center}
\end{table}

This so-called mass-dependent shift has been calculated from four different mass measurements: $^{6}$Li \cite{Bro10}, $^{23}$Na, $^{39}$K, and $^{41}$K using as calibrant $^{7}$Li, H$_{3}$O, $^{23}$Na, and H$_{3}$O, respectively.
\begin{figure}[ht]
  \begin{center}{
    \includegraphics[width=0.7\textwidth]{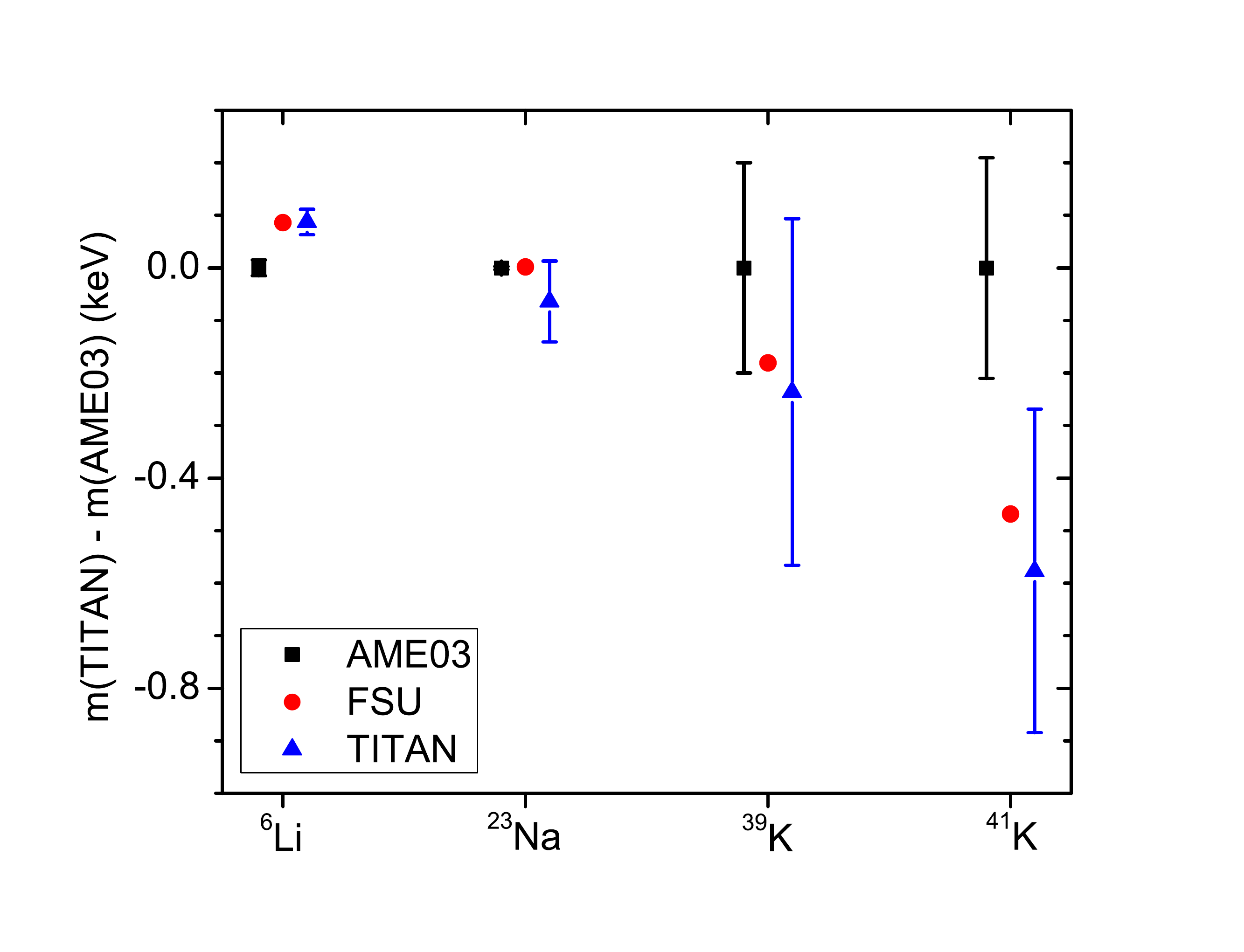}}
    \caption{Masses of $^{6}$Li \cite{Bro10}, $^{23}$Na, $^{39}$K, and $^{41}$K measured by the TITAN Penning trap using $^{7}$Li, H$_{3}$O,  $^{23}$Na and H$_{3}$O respectively as calibrant, compared to both the AME03 \cite{Aud03} and a more recent measurement from the FSU Penning trap \cite{Mou10}.}
    {\label{fig:Na23K39K41}}
  \end{center}
\end{figure}
These measurements were taken using the TITAN off-line ion source and a trapping potential of $V_{0}$ = 35.7V.
The resulting frequency ratios are presented in table~\ref{tab:MassDepShift}, while the mass difference for $^{6}$Li, $^{23}$Na, $^{39}$K and $^{41}$K compared to the 2003 Atomic Mass Evaluation (AME03) \cite{Aud03} values and a more precise measurement from the Florida State University (FSU) Penning trap \cite{Mou10} are shown in figure~\ref{fig:Na23K39K41}. All masses agree with the FSU measurement within one $\sigma$. The total shift in the frequency ratio was taken as the weighted mean of the three measurements yielding -0.2(2) ppb/u. By dividing this value by the trapping voltage used: $V_{0}$ = 35.7V, one obtains a relative change in the cyclotron frequency ratio due to the total combined systematic effects equal to
\begin{equation}\label{eq:44b}
(\Delta R/R)_{total} = -4(6) \times 10^{-12}  \cdot V_{0} \cdot \Delta (m/q). 
\end{equation}
Note that if the less precise AME03 masses are used instead, the mass-dependant shift becomes -0.6(3) ppb/u and \((\Delta R/R)_{total} = -1.7(8) \times 10^{-11}  \cdot V_{0} \cdot \Delta (m/q))\). 

\section{Summary and outlook} \label{Sec:7}

Penning trap mass measurements to a level of precision and accuracy of $\delta m/m \sim$10$^{-9}$ are only made possible if detailed systematic studies of the system is performed. Such studies of the TITAN Penning trap are presented, in particular the different sources of systematic errors on the measured cyclotron frequency arising from the imperfections of the Penning trap, such as the magnetic field inhomogeneities, the misalignment of the trap electrodes with the magnetic field, the harmonic distortion of the trap potential, and the non-harmonic terms in the trapping potential. 

The total systematic error on frequency ratio determination at TITAN was found to be \((\Delta R/R)_{total} = -4(6) \times 10^{-12} \cdot \Delta (m/q) \cdot V_{0}\). These mass-dependant systematic error depend on the mass-to-charge ratio difference $\Delta (m/q)$. This means that when the mass measurement is performed using a calibrant and species of similar mass-to-charge ratio, the shift on the frequency ratio will be effectively quenched. Also, these estimates are proportional to the trapping potential $V_{0}$, and their contribution to the mass measurement systematic error can be reduced by using a small trapping potential, hence operating in a so-called shallow trap.

We also presented compensation of the trapping potential using a new general method of compensation. We performed this compensation using two different methods in order to optimize the combination of correction tube and guard voltage that provide the maximal compensation of the trapping potential. Based on this, the TITAN Penning trap is able to perform accurate mass measurements at a level of precision of below one ppb.

\section{Acknowledgements} \label{Sec:8}

This work was supported by the Natural Sciences and Engineering
Research Council of Canada (NSERC) and the National Research Council
of Canada (NRC). S.E.~acknowledges support from the Vanier
CGS program, T.B.~from the Evangelisches Studienwerk e.V.~Villigst, 
A.G.~from the NSERC PGS-M program and D.L.~from TRIUMF during his 2007-2008 sabbatical.





\bibliographystyle{model1c-num-names}
\bibliography{<your-bib-database>}



\end{document}